\def\DIRvalue{Hosomichi}
\def\IDvalue{HO}
\def\titlevalue{${\cal N}=2$ SUSY gauge theories on $\bf S^4$}
\def\authorvalue{Kazuo Hosomichi}
\def\shortauthorvalue{\authorvalue}
\def\addressvalue{Department of Physics, National Taiwan University,
  Taipei 10617, Taiwan\\
\tt hosomiti@phys.ntu.edu.tw}
\def\abstractvalue{We review exact results in ${\cal N}=2$
supersymmetric gauge theories defined on $S^4$ and its deformation. We
first summarize the construction of rigid SUSY theories on curved
backgrounds based on off-shell supergravity, then explain how to apply
localization principle to supersymmetric path integrals. Closed formulae
for partition function as well as expectation values of non-local BPS
observables are presented.}
\def\preprintvalue{}

\ifx\ifLONG\undefined
\documentclass[11pt]{article}
\newcommand{\chapterauthor}[1]{
\begin{center}
{\bf \normalsize  #1}
\end{center}
}

\newcommand{\chapteraddress}[1]{
\begin{center}
{ \small \it \addressvalue}
\end{center}
}

\newcommand{\chapterabstract}[1]{
\vspace{\baselineskip}
\begin{center}
\textbf{\small Abstract}
\end{center}
#1}

\newcommand{\chapterheader}{

\chapter[\titlevalue{}  (by \shortauthorvalue)]{\titlevalue}
\label{Chapter\IDvalue}
\chapterauthor{\authorvalue}
\chapteraddress{\addressvalue}
\chapterabstract{\abstractvalue}
\tightmtctrue
\minitoc
}

\newcommand{\documentheader}{
\begin{flushright} \small
  \preprintvalue
 \end{flushright}

\begin{center}
{\bf \Large \titlevalue}
\end{center}

\chapterauthor{\authorvalue}
\chapteraddress{\addressvalue}
\chapterabstract{\abstractvalue}

\medskip

This is a contribution to the review volume ``Localization techniques
in quantum field theories'' (eds. V.~Pestun and M.~Zabzine) which
contains 17 Chapters available at \cite{ContributionSummary}

\tableofcontents
}

\newcommand{\ifvolume}[2]{\ifx\ifLONG\undefined#2\else#1\fi}

\newcommand{\documentfinish}{
\ifx\ifLONG\undefined
\bibliographystyle{bibreview} 
\bibliography{\IDvalue,review}  
\end{document}
\else
\addcontentsline{toc}{section}{References}
\providecommand{\href}[2]{#2}\begingroup\raggedright\begin{thebibliography}{10}

\bibitem{ContributionSummary}
V.~Pestun and M.~Zabzine, eds., {\em Localization techniques in quantum field
  theory}, vol.~xx.
\newblock Journal of Physics A, 2016.
\newblock \href{http://arxiv.org/abs/1608.02952}{{\tt 1608.02952}}.
\newblock \url{https://arxiv.org/src/1608.02952/anc/LocQFT.pdf},
  \url{http://pestun.ihes.fr/pages/LocalizationReview/LocQFT.pdf}.

\bibitem{HO:Seiberg:1994rs}
N.~Seiberg and E.~Witten, ``{Electric - magnetic duality, monopole
  condensation, and confinement in ${\cal N}=2$ supersymmetric Yang-Mills
  theory},'' \href{http://dx.doi.org/10.1016/0550-3213(94)90124-4}{{\em Nucl.
  Phys.} {\bf B426} (1994)  19--52},
  \href{http://arxiv.org/abs/hep-th/9407087}{{\tt arXiv:hep-th/9407087
  [hep-th]}}.
[Erratum: Nucl. Phys.B430,485(1994)].

\bibitem{HO:Seiberg:1994aj}
N.~Seiberg and E.~Witten, ``{Monopoles, duality and chiral symmetry breaking in
  ${\cal N}=2$ supersymmetric QCD},''
  \href{http://dx.doi.org/10.1016/0550-3213(94)90214-3}{{\em Nucl. Phys.} {\bf
  B431} (1994)  484--550},
\href{http://arxiv.org/abs/hep-th/9408099}{{\tt arXiv:hep-th/9408099
  [hep-th]}}.

\bibitem{HO:Witten:1988ze}
E.~Witten, ``{Topological Quantum Field Theory},''
\href{http://dx.doi.org/10.1007/BF01223371}{{\em Commun. Math. Phys.} {\bf 117}
  (1988)  353}.

\bibitem{HO:Nekrasov:2002qd}
N.~A. Nekrasov, ``{Seiberg-Witten prepotential from instanton counting},''
  \href{http://dx.doi.org/10.4310/ATMP.2003.v7.n5.a4}{{\em Adv. Theor. Math.
  Phys.} {\bf 7} (2003) no.~5, 831--864},
\href{http://arxiv.org/abs/hep-th/0206161}{{\tt arXiv:hep-th/0206161
  [hep-th]}}.

\bibitem{HO:Losev:1997tp}
A.~Losev, N.~Nekrasov, and S.~L. Shatashvili, ``{Issues in topological gauge
  theory},'' \href{http://dx.doi.org/10.1016/S0550-3213(98)00628-2}{{\em Nucl.
  Phys.} {\bf B534} (1998)  549--611},
\href{http://arxiv.org/abs/hep-th/9711108}{{\tt arXiv:hep-th/9711108
  [hep-th]}}.

\bibitem{HO:Moore:1997dj}
G.~W. Moore, N.~Nekrasov, and S.~Shatashvili, ``{Integrating over Higgs
  branches},'' \href{http://dx.doi.org/10.1007/PL00005525}{{\em Commun. Math.
  Phys.} {\bf 209} (2000)  97--121},
\href{http://arxiv.org/abs/hep-th/9712241}{{\tt arXiv:hep-th/9712241
  [hep-th]}}.

\bibitem{HO:Lossev:1997bz}
A.~Lossev, N.~Nekrasov, and S.~L. Shatashvili, ``{Testing Seiberg-Witten
  solution},'' in {\em {Strings, branes and dualities. Proceedings, NATO
  Advanced Study Institute, Cargese, France, May 26-June 14, 1997}}.
\newblock 1997.
\newblock
\href{http://arxiv.org/abs/hep-th/9801061}{{\tt arXiv:hep-th/9801061
  [hep-th]}}.
\newblock

\bibitem{HO:Pestun:2007rz}
V.~Pestun, ``{Localization of gauge theory on a four-sphere and supersymmetric
  Wilson loops},'' \href{http://dx.doi.org/10.1007/s00220-012-1485-0}{{\em
  Commun. Math. Phys.} {\bf 313} (2012)  71--129},
\href{http://arxiv.org/abs/0712.2824}{{\tt arXiv:0712.2824 [hep-th]}}.

\bibitem{HO:Erickson:2000af}
J.~K. Erickson, G.~W. Semenoff, and K.~Zarembo, ``{Wilson loops in ${\cal N}=4$
  supersymmetric Yang-Mills theory},''
  \href{http://dx.doi.org/10.1016/S0550-3213(00)00300-X}{{\em Nucl. Phys.} {\bf
  B582} (2000)  155--175},
\href{http://arxiv.org/abs/hep-th/0003055}{{\tt arXiv:hep-th/0003055
  [hep-th]}}.

\bibitem{HO:Drukker:2000rr}
N.~Drukker and D.~J. Gross, ``{An Exact prediction of ${\cal N}=4$ SUSYM theory
  for string theory},'' \href{http://dx.doi.org/10.1063/1.1372177}{{\em J.
  Math. Phys.} {\bf 42} (2001)  2896--2914},
\href{http://arxiv.org/abs/hep-th/0010274}{{\tt arXiv:hep-th/0010274
  [hep-th]}}.

\bibitem{HO:Okuda:2010ke}
T.~Okuda and V.~Pestun, ``{On the instantons and the hypermultiplet mass of
  ${\cal N}=2^\ast$ super Yang-Mills on $S^4$},''
  \href{http://dx.doi.org/10.1007/JHEP03(2012)017}{{\em JHEP} {\bf 03} (2012)
  017},
\href{http://arxiv.org/abs/1004.1222}{{\tt arXiv:1004.1222 [hep-th]}}.

\bibitem{ContributionZA}
K.~Zarembo, ``Localization and AdS/CFT Correspondence,'' {\em Journal of
  Physics A} {\bf xx} (2016)  000, \href{http://arxiv.org/abs/1608.02963}{{\tt
  1608.02963}}.

\bibitem{HO:Kapustin:2009kz}
A.~Kapustin, B.~Willett, and I.~Yaakov, ``{Exact Results for Wilson Loops in
  Superconformal Chern-Simons Theories with Matter},''
  \href{http://dx.doi.org/10.1007/JHEP03(2010)089}{{\em JHEP} {\bf 03} (2010)
  089},
\href{http://arxiv.org/abs/0909.4559}{{\tt arXiv:0909.4559 [hep-th]}}.

\bibitem{HO:Jafferis:2010un}
D.~L. Jafferis, ``{The Exact Superconformal R-Symmetry Extremizes $Z$},''
  \href{http://dx.doi.org/10.1007/JHEP05(2012)159}{{\em JHEP} {\bf 05} (2012)
  159},
\href{http://arxiv.org/abs/1012.3210}{{\tt arXiv:1012.3210 [hep-th]}}.

\bibitem{HO:Hama:2010av}
N.~Hama, K.~Hosomichi, and S.~Lee, ``{Notes on SUSY Gauge Theories on
  Three-Sphere},'' \href{http://dx.doi.org/10.1007/JHEP03(2011)127}{{\em JHEP}
  {\bf 03} (2011)  127},
\href{http://arxiv.org/abs/1012.3512}{{\tt arXiv:1012.3512 [hep-th]}}.

\bibitem{HO:Benini:2012ui}
F.~Benini and S.~Cremonesi, ``{Partition Functions of ${\mathcal{N}=(2,2)}$
  Gauge Theories on $S^2$ and Vortices},''
  \href{http://dx.doi.org/10.1007/s00220-014-2112-z}{{\em Commun. Math. Phys.}
  {\bf 334} (2015) no.~3, 1483--1527},
\href{http://arxiv.org/abs/1206.2356}{{\tt arXiv:1206.2356 [hep-th]}}.

\bibitem{HO:Doroud:2012xw}
N.~Doroud, J.~Gomis, B.~Le~Floch, and S.~Lee, ``{Exact Results in $D=2$
  Supersymmetric Gauge Theories},''
  \href{http://dx.doi.org/10.1007/JHEP05(2013)093}{{\em JHEP} {\bf 05} (2013)
  093},
\href{http://arxiv.org/abs/1206.2606}{{\tt arXiv:1206.2606 [hep-th]}}.

\bibitem{HO:Kallen:2012va}
J.~{K\"all\'en}, J.~Qiu, and M.~Zabzine, ``{The perturbative partition function
  of supersymmetric 5D Yang-Mills theory with matter on the five-sphere},''
  \href{http://dx.doi.org/10.1007/JHEP08(2012)157}{{\em JHEP} {\bf 08} (2012)
  157},
\href{http://arxiv.org/abs/1206.6008}{{\tt arXiv:1206.6008 [hep-th]}}.

\bibitem{HO:Kim:2012ava}
H.-C. Kim and S.~Kim, ``{M5-branes from gauge theories on the 5-sphere},''
  \href{http://dx.doi.org/10.1007/JHEP05(2013)144}{{\em JHEP} {\bf 05} (2013)
  144},
\href{http://arxiv.org/abs/1206.6339}{{\tt arXiv:1206.6339 [hep-th]}}.

\bibitem{HO:Kim:2012qf}
H.-C. Kim, J.~Kim, and S.~Kim, ``{Instantons on the 5-sphere and M5-branes},''
\href{http://arxiv.org/abs/1211.0144}{{\tt arXiv:1211.0144 [hep-th]}}.

\bibitem{HO:Gerchkovitz:2014gta}
E.~Gerchkovitz, J.~Gomis, and Z.~Komargodski, ``{Sphere Partition Functions and
  the Zamolodchikov Metric},''
  \href{http://dx.doi.org/10.1007/JHEP11(2014)001}{{\em JHEP} {\bf 11} (2014)
  001},
\href{http://arxiv.org/abs/1405.7271}{{\tt arXiv:1405.7271 [hep-th]}}.

\bibitem{HO:Gomis:2014woa}
J.~Gomis and N.~Ishtiaque, ``{K\"ahler potential and ambiguities in 4d $
  \mathcal{N} $ = 2 SCFTs},''
  \href{http://dx.doi.org/10.1007/JHEP04(2015)169}{{\em JHEP} {\bf 04} (2015)
  169}, \href{http://arxiv.org/abs/1409.5325}{{\tt arXiv:1409.5325 [hep-th]}}.
[JHEP04,169(2015)].

\bibitem{HO:Gaiotto:2009we}
D.~Gaiotto, ``{${\cal N}=2$ dualities},''
  \href{http://dx.doi.org/10.1007/JHEP08(2012)034}{{\em JHEP} {\bf 08} (2012)
  034},
\href{http://arxiv.org/abs/0904.2715}{{\tt arXiv:0904.2715 [hep-th]}}.

\bibitem{HO:Alday:2009aq}
L.~F. Alday, D.~Gaiotto, and Y.~Tachikawa, ``{Liouville Correlation Functions
  from Four-dimensional Gauge Theories},''
  \href{http://dx.doi.org/10.1007/s11005-010-0369-5}{{\em Lett. Math. Phys.}
  {\bf 91} (2010)  167--197},
\href{http://arxiv.org/abs/0906.3219}{{\tt arXiv:0906.3219 [hep-th]}}.

\bibitem{ContributionTA}
Y.~Tachikawa, ``A brief review of the 2d/4d correspondences,'' {\em Journal of
  Physics A} {\bf xx} (2016)  000, \href{http://arxiv.org/abs/1608.02964}{{\tt
  1608.02964}}.

\bibitem{HO:Wyllard:2009hg}
N.~Wyllard, ``{$A_{N-1}$ conformal Toda field theory correlation functions from
  conformal ${\cal N}=2$ $SU(N)$ quiver gauge theories},''
  \href{http://dx.doi.org/10.1088/1126-6708/2009/11/002}{{\em JHEP} {\bf 11}
  (2009)  002},
\href{http://arxiv.org/abs/0907.2189}{{\tt arXiv:0907.2189 [hep-th]}}.

\bibitem{HO:Hama:2011ea}
N.~Hama, K.~Hosomichi, and S.~Lee, ``{SUSY Gauge Theories on Squashed
  Three-Spheres},'' \href{http://dx.doi.org/10.1007/JHEP05(2011)014}{{\em JHEP}
  {\bf 05} (2011)  014},
\href{http://arxiv.org/abs/1102.4716}{{\tt arXiv:1102.4716 [hep-th]}}.

\bibitem{HO:Imamura:2011wg}
Y.~Imamura and D.~Yokoyama, ``{${\cal N}=2$ supersymmetric theories on squashed
  three-sphere},'' \href{http://dx.doi.org/10.1103/PhysRevD.85.025015}{{\em
  Phys. Rev.} {\bf D85} (2012)  025015},
\href{http://arxiv.org/abs/1109.4734}{{\tt arXiv:1109.4734 [hep-th]}}.

\bibitem{HO:Festuccia:2011ws}
G.~Festuccia and N.~Seiberg, ``{Rigid Supersymmetric Theories in Curved
  Superspace},'' \href{http://dx.doi.org/10.1007/JHEP06(2011)114}{{\em JHEP}
  {\bf 06} (2011)  114},
\href{http://arxiv.org/abs/1105.0689}{{\tt arXiv:1105.0689 [hep-th]}}.

\bibitem{ContributionDU}
T.~Dumitrescu, ``An Introduction to Supersymmetric Field Theories in Curved
  Space,'' {\em Journal of Physics A} {\bf xx} (2016)  000,
  \href{http://arxiv.org/abs/1608.02957}{{\tt 1608.02957}}.

\bibitem{HO:Hama:2012bg}
N.~Hama and K.~Hosomichi, ``{Seiberg-Witten Theories on Ellipsoids},''
  \href{http://dx.doi.org/10.1007/JHEP09(2012)033,
  10.1007/JHEP10(2012)051}{{\em JHEP} {\bf 09} (2012)  033},
  \href{http://arxiv.org/abs/1206.6359}{{\tt arXiv:1206.6359 [hep-th]}}.
[Addendum: JHEP10,051(2012)].

\bibitem{HO:Samtleben:2012gy}
H.~Samtleben and D.~Tsimpis, ``{Rigid supersymmetric theories in 4d Riemannian
  space},'' \href{http://dx.doi.org/10.1007/JHEP05(2012)132}{{\em JHEP} {\bf
  05} (2012)  132},
\href{http://arxiv.org/abs/1203.3420}{{\tt arXiv:1203.3420 [hep-th]}}.

\bibitem{HO:Dumitrescu:2012ha}
T.~T. Dumitrescu, G.~Festuccia, and N.~Seiberg, ``{Exploring Curved
  Superspace},'' \href{http://dx.doi.org/10.1007/JHEP08(2012)141}{{\em JHEP}
  {\bf 08} (2012)  141},
\href{http://arxiv.org/abs/1205.1115}{{\tt arXiv:1205.1115 [hep-th]}}.

\bibitem{HO:Cassani:2012ri}
D.~Cassani, C.~Klare, D.~Martelli, A.~Tomasiello, and A.~Zaffaroni,
  ``{Supersymmetry in Lorentzian Curved Spaces and Holography},''
  \href{http://dx.doi.org/10.1007/s00220-014-1983-3}{{\em Commun. Math. Phys.}
  {\bf 327} (2014)  577--602},
\href{http://arxiv.org/abs/1207.2181}{{\tt arXiv:1207.2181 [hep-th]}}.

\bibitem{HO:Liu:2012bi}
J.~T. Liu, L.~A. Pando~Zayas, and D.~Reichmann, ``{Rigid Supersymmetric
  Backgrounds of Minimal Off-Shell Supergravity},''
  \href{http://dx.doi.org/10.1007/JHEP10(2012)034}{{\em JHEP} {\bf 10} (2012)
  034},
\href{http://arxiv.org/abs/1207.2785}{{\tt arXiv:1207.2785 [hep-th]}}.

\bibitem{HO:Dumitrescu:2012at}
T.~T. Dumitrescu and G.~Festuccia, ``{Exploring Curved Superspace (II)},''
  \href{http://dx.doi.org/10.1007/JHEP01(2013)072}{{\em JHEP} {\bf 01} (2013)
  072},
\href{http://arxiv.org/abs/1209.5408}{{\tt arXiv:1209.5408 [hep-th]}}.

\bibitem{HO:Klare:2013dka}
C.~Klare and A.~Zaffaroni, ``{Extended Supersymmetry on Curved Spaces},''
  \href{http://dx.doi.org/10.1007/JHEP10(2013)218}{{\em JHEP} {\bf 10} (2013)
  218},
\href{http://arxiv.org/abs/1308.1102}{{\tt arXiv:1308.1102 [hep-th]}}.

\bibitem{HO:Nosaka:2013cpa}
T.~Nosaka and S.~Terashima, ``{Supersymmetric Gauge Theories on a Squashed
  Four-Sphere},'' \href{http://dx.doi.org/10.1007/JHEP12(2013)001}{{\em JHEP}
  {\bf 12} (2013)  001},
\href{http://arxiv.org/abs/1310.5939}{{\tt arXiv:1310.5939 [hep-th]}}.

\bibitem{HO:Cabo-Bizet:2014nia}
A.~Cabo-Bizet, E.~Gava, V.~I. Giraldo-Rivera, M.~N. Muteeb, and K.~S. Narain,
  ``{Partition Function of $N=2$ Gauge Theories on a Squashed $S^4$ with
  $SU(2)\times U(1)$ Isometry},''
  \href{http://dx.doi.org/10.1016/j.nuclphysb.2015.07.029}{{\em Nucl. Phys.}
  {\bf B899} (2015)  149--164},
\href{http://arxiv.org/abs/1412.6826}{{\tt arXiv:1412.6826 [hep-th]}}.

\bibitem{HO:Bawane:2014uka}
A.~Bawane, G.~Bonelli, M.~Ronzani, and A.~Tanzini, ``{$\mathcal{N}=2$
  supersymmetric gauge theories on $S^2\times S^2$ and Liouville Gravity},''
  \href{http://dx.doi.org/10.1007/JHEP07(2015)054}{{\em JHEP} {\bf 07} (2015)
  054},
\href{http://arxiv.org/abs/1411.2762}{{\tt arXiv:1411.2762 [hep-th]}}.

\bibitem{HO:Sinamuli:2014lma}
M.~Sinamuli, ``{On ${\cal N}=2$ supersymmetric gauge theories on $S^2\times
  S^2$},''
\href{http://arxiv.org/abs/1411.4918}{{\tt arXiv:1411.4918 [hep-th]}}.

\bibitem{HO:Murthy:2015yfa}
S.~Murthy and V.~Reys, ``{Functional determinants, index theorems, and exact
  quantum black hole entropy},''
  \href{http://dx.doi.org/10.1007/JHEP12(2015)028}{{\em JHEP} {\bf 12} (2015)
  028},
\href{http://arxiv.org/abs/1504.01400}{{\tt arXiv:1504.01400 [hep-th]}}.

\bibitem{HO:Gupta:2015gga}
R.~K. Gupta, Y.~Ito, and I.~Jeon, ``{Supersymmetric Localization for BPS Black
  Hole Entropy: 1-loop Partition Function from Vector Multiplets},''
  \href{http://dx.doi.org/10.1007/JHEP11(2015)197}{{\em JHEP} {\bf 11} (2015)
  197},
\href{http://arxiv.org/abs/1504.01700}{{\tt arXiv:1504.01700 [hep-th]}}.

\bibitem{HO:Butter:2015tra}
D.~Butter, G.~Inverso, and I.~Lodato, ``{Rigid 4D $\mathcal{N}=2$
  supersymmetric backgrounds and actions},''
  \href{http://dx.doi.org/10.1007/JHEP09(2015)088}{{\em JHEP} {\bf 09} (2015)
  088},
\href{http://arxiv.org/abs/1505.03500}{{\tt arXiv:1505.03500 [hep-th]}}.

\bibitem{HO:Nishioka:2013haa}
T.~Nishioka and I.~Yaakov, ``{Supersymmetric R\'enyi Entropy},''
  \href{http://dx.doi.org/10.1007/JHEP10(2013)155}{{\em JHEP} {\bf 10} (2013)
  155},
\href{http://arxiv.org/abs/1306.2958}{{\tt arXiv:1306.2958 [hep-th]}}.

\bibitem{HO:Huang:2014pda}
X.~Huang and Y.~Zhou, ``{$\mathcal{N}=4 $ Super-Yang-Mills on conic space as
  hologram of STU topological black hole},''
  \href{http://dx.doi.org/10.1007/JHEP02(2015)068}{{\em JHEP} {\bf 02} (2015)
  068},
\href{http://arxiv.org/abs/1408.3393}{{\tt arXiv:1408.3393 [hep-th]}}.

\bibitem{HO:Crossley:2014oea}
M.~Crossley, E.~Dyer, and J.~Sonner, ``{Super-R\'enyi entropy \& Wilson loops
  for $\mathcal{N}=4$ SYM and their gravity duals},''
  \href{http://dx.doi.org/10.1007/JHEP12(2014)001}{{\em JHEP} {\bf 12} (2014)
  001},
\href{http://arxiv.org/abs/1409.0542}{{\tt arXiv:1409.0542 [hep-th]}}.

\bibitem{HO:Hama:2014iea}
N.~Hama, T.~Nishioka, and T.~Ugajin, ``{Supersymmetric R\'enyi entropy in five
  dimensions},'' \href{http://dx.doi.org/10.1007/JHEP12(2014)048}{{\em JHEP}
  {\bf 12} (2014)  048},
\href{http://arxiv.org/abs/1410.2206}{{\tt arXiv:1410.2206 [hep-th]}}.

\bibitem{HO:Zhou:2015cpa}
Y.~Zhou, ``{Universal Features of Four-Dimensional Superconformal Field Theory
  on Conic Space},'' \href{http://dx.doi.org/10.1007/JHEP08(2015)052}{{\em
  JHEP} {\bf 08} (2015)  052},
\href{http://arxiv.org/abs/1506.06512}{{\tt arXiv:1506.06512 [hep-th]}}.

\bibitem{HO:Alday:2009fs}
L.~F. Alday, D.~Gaiotto, S.~Gukov, Y.~Tachikawa, and H.~Verlinde, ``{Loop and
  surface operators in ${\cal N}=2$ gauge theory and Liouville modular
  geometry},'' \href{http://dx.doi.org/10.1007/JHEP01(2010)113}{{\em JHEP} {\bf
  01} (2010)  113},
\href{http://arxiv.org/abs/0909.0945}{{\tt arXiv:0909.0945 [hep-th]}}.

\bibitem{HO:Drukker:2009id}
N.~Drukker, J.~Gomis, T.~Okuda, and J.~Teschner, ``{Gauge Theory Loop Operators
  and Liouville Theory},''
  \href{http://dx.doi.org/10.1007/JHEP02(2010)057}{{\em JHEP} {\bf 02} (2010)
  057},
\href{http://arxiv.org/abs/0909.1105}{{\tt arXiv:0909.1105 [hep-th]}}.

\bibitem{HO:Gomis:2011pf}
J.~Gomis, T.~Okuda, and V.~Pestun, ``{Exact Results for 't Hooft Loops in Gauge
  Theories on $S^4$},'' \href{http://dx.doi.org/10.1007/JHEP05(2012)141}{{\em
  JHEP} {\bf 05} (2012)  141},
\href{http://arxiv.org/abs/1105.2568}{{\tt arXiv:1105.2568 [hep-th]}}.

\bibitem{HO:Fucito:2015ofa}
F.~Fucito, J.~F. Morales, and R.~Poghossian, ``{Wilson loops and chiral
  correlators on squashed spheres},''
  \href{http://dx.doi.org/10.1007/JHEP11(2015)064}{{\em JHEP} {\bf 11} (2015)
  064},
\href{http://arxiv.org/abs/1507.05426}{{\tt arXiv:1507.05426 [hep-th]}}.

\bibitem{HO:Gukov:2014gja}
S.~Gukov, \href{http://dx.doi.org/10.1007/978-3-319-18769-3_8}{``{Surface
  Operators},''} in {\em New Dualities of Supersymmetric Gauge Theories},
  J.~Teschner, ed., pp.~223--259.
\newblock Springer International Publishing, 2016.
\newblock
\href{http://arxiv.org/abs/1412.7127}{{\tt arXiv:1412.7127 [hep-th]}}.
\newblock

\bibitem{HO:Gukov:2006jk}
S.~Gukov and E.~Witten, ``{Gauge Theory, Ramification, And The Geometric
  Langlands Program},''
\href{http://arxiv.org/abs/hep-th/0612073}{{\tt arXiv:hep-th/0612073
  [hep-th]}}.

\bibitem{HO:Alday:2010vg}
L.~F. Alday and Y.~Tachikawa, ``{Affine $SL(2)$ conformal blocks from 4d gauge
  theories},'' \href{http://dx.doi.org/10.1007/s11005-010-0422-4}{{\em Lett.
  Math. Phys.} {\bf 94} (2010)  87--114},
\href{http://arxiv.org/abs/1005.4469}{{\tt arXiv:1005.4469 [hep-th]}}.

\bibitem{HO:Dimofte:2010tz}
T.~Dimofte, S.~Gukov, and L.~Hollands, ``{Vortex Counting and Lagrangian
  3-manifolds},'' \href{http://dx.doi.org/10.1007/s11005-011-0531-8}{{\em Lett.
  Math. Phys.} {\bf 98} (2011)  225--287},
\href{http://arxiv.org/abs/1006.0977}{{\tt arXiv:1006.0977 [hep-th]}}.

\bibitem{HO:Awata:2010bz}
H.~Awata, H.~Fuji, H.~Kanno, M.~Manabe, and Y.~Yamada, ``{Localization with a
  Surface Operator, Irregular Conformal Blocks and Open Topological String},''
  \href{http://dx.doi.org/10.4310/ATMP.2012.v16.n3.a1}{{\em Adv. Theor. Math.
  Phys.} {\bf 16} (2012) no.~3, 725--804},
\href{http://arxiv.org/abs/1008.0574}{{\tt arXiv:1008.0574 [hep-th]}}.

\bibitem{HO:Kozcaz:2010yp}
C.~Kozcaz, S.~Pasquetti, F.~Passerini, and N.~Wyllard, ``{Affine $sl(N)$
  conformal blocks from ${\cal N}=2$ $SU(N)$ gauge theories},''
  \href{http://dx.doi.org/10.1007/JHEP01(2011)045}{{\em JHEP} {\bf 01} (2011)
  045},
\href{http://arxiv.org/abs/1008.1412}{{\tt arXiv:1008.1412 [hep-th]}}.

\bibitem{HO:Wyllard:2010rp}
N.~Wyllard, ``{W-algebras and surface operators in ${\cal N}=2$ gauge
  theories},'' \href{http://dx.doi.org/10.1088/1751-8113/44/15/155401}{{\em J.
  Phys.} {\bf A44} (2011)  155401},
\href{http://arxiv.org/abs/1011.0289}{{\tt arXiv:1011.0289 [hep-th]}}.

\bibitem{HO:Wyllard:2010vi}
N.~Wyllard, ``{Instanton partition functions in ${\cal N}=2$ $SU(N)$ gauge
  theories with a general surface operator, and their W-algebra duals},''
  \href{http://dx.doi.org/10.1007/JHEP02(2011)114}{{\em JHEP} {\bf 02} (2011)
  114},
\href{http://arxiv.org/abs/1012.1355}{{\tt arXiv:1012.1355 [hep-th]}}.

\bibitem{HO:Kanno:2011fw}
H.~Kanno and Y.~Tachikawa, ``{Instanton counting with a surface operator and
  the chain-saw quiver},''
  \href{http://dx.doi.org/10.1007/JHEP06(2011)119}{{\em JHEP} {\bf 06} (2011)
  119},
\href{http://arxiv.org/abs/1105.0357}{{\tt arXiv:1105.0357 [hep-th]}}.

\bibitem{HO:Gomis:2014eya}
J.~Gomis and B.~Le~Floch, ``{M2-brane surface operators and gauge theory
  dualities in Toda},''
\href{http://arxiv.org/abs/1407.1852}{{\tt arXiv:1407.1852 [hep-th]}}.

\bibitem{HO:Nawata:2014nca}
S.~Nawata, ``{Givental $J$-functions, Quantum integrable systems, AGT relation
  with surface operator},''
\href{http://arxiv.org/abs/1408.4132}{{\tt arXiv:1408.4132 [hep-th]}}.

\bibitem{HO:Wess:1992cp}
J.~Wess and J.~Bagger, {\em {Supersymmetry and supergravity}}.
\newblock Princeton Univ. Pr.,
1992.
\newblock

\bibitem{HO:deWit:1979dzm}
B.~de~Wit, J.~W. van Holten, and A.~Van~Proeyen, ``{Transformation Rules of
  ${\cal N}=2$ Supergravity Multiplets},''
\href{http://dx.doi.org/10.1016/0550-3213(80)90125-X}{{\em Nucl. Phys.} {\bf
  B167} (1980)  186}.

\bibitem{HO:deWit:1980lyi}
B.~de~Wit, J.~W. van Holten, and A.~Van~Proeyen, ``{Structure of ${\cal N}=2$
  Supergravity},'' \href{http://dx.doi.org/10.1016/0550-3213(81)90211-X}{{\em
  Nucl. Phys.} {\bf B184} (1981)  77}.
[Erratum: Nucl. Phys.B222,516(1983)].

\bibitem{HO:deWit:1983xe}
B.~de~Wit, P.~G. Lauwers, R.~Philippe, and A.~Van~Proeyen, ``{Noncompact ${\cal
  N}=2$ Supergravity},''
\href{http://dx.doi.org/10.1016/0370-2693(84)90395-2}{{\em Phys. Lett.} {\bf
  B135} (1984)  295}.

\bibitem{HO:deWit:1984rvr}
B.~de~Wit, P.~G. Lauwers, and A.~Van~Proeyen, ``{Lagrangians of ${\cal N}=2$
  Supergravity - Matter Systems},''
\href{http://dx.doi.org/10.1016/0550-3213(85)90154-3}{{\em Nucl. Phys.} {\bf
  B255} (1985)  569}.

\bibitem{HO:Mohaupt:2000mj}
T.~Mohaupt, ``{Black hole entropy, special geometry and strings},'' {\em
  Fortsch. Phys.} {\bf 49} (2001)  3--161,
\href{http://arxiv.org/abs/hep-th/0007195}{{\tt arXiv:hep-th/0007195
  [hep-th]}}.

\bibitem{HO:Freedman:2012zz}
D.~Z. Freedman and A.~Van~Proeyen, {\em {Supergravity}}.
\newblock Cambridge Univ. Press, Cambridge, UK,
2012.
\newblock

\bibitem{HO:Berkovits:1993hx}
N.~Berkovits, ``{A Ten-dimensional superYang-Mills action with off-shell
  supersymmetry},'' \href{http://dx.doi.org/10.1016/0370-2693(93)91791-K}{{\em
  Phys. Lett.} {\bf B318} (1993)  104--106},
\href{http://arxiv.org/abs/hep-th/9308128}{{\tt arXiv:hep-th/9308128
  [hep-th]}}.

\bibitem{HO:Pestun:2014mja}
V.~Pestun, \href{http://dx.doi.org/10.1007/978-3-319-18769-3_6}{``{Localization
  for $\mathcal {N}=2$ Supersymmetric Gauge Theories in Four Dimensions},''} in
  {\em New Dualities of Supersymmetric Gauge Theories}, J.~Teschner, ed.,
  pp.~159--194.
\newblock Springer International Publishing, 2016.
\newblock
\href{http://arxiv.org/abs/1412.7134}{{\tt arXiv:1412.7134 [hep-th]}}.
\newblock

\bibitem{HO:Witten:1982im}
E.~Witten, ``{Supersymmetry and Morse theory},''
{\em J. Diff. Geom.} {\bf 17} (1982) no.~4, 661--692.

\bibitem{HO:Witten:1984aa}
E.~Witten, ``{Holomorphic Morse inequalities},'' in {\em Algebraic and
  differential topology, global differential geometry}, G.~M. Rassias, ed.,
  Teubner-Texte zur Mathematik, pp.~318--333.
\newblock Teubner, 1984.

\bibitem{HO:Atiyah:1974aa}
M.~F. Atiyah, {\em {Elliptic operators and compact groups}}.
\newblock Springer-Verlag Berlin, 1974.

\bibitem{HO:Zamolodchikov:1995aa}
A.~B. Zamolodchikov and A.~B. Zamolodchikov, ``{Structure constants and
  conformal bootstrap in Liouville field theory},''
  \href{http://dx.doi.org/10.1016/0550-3213(96)00351-3}{{\em Nucl. Phys.} {\bf
  B477} (1996)  577--605},
\href{http://arxiv.org/abs/hep-th/9506136}{{\tt arXiv:hep-th/9506136
  [hep-th]}}.

\bibitem{HO:'tHooft:1977hy}
G.~'t~Hooft, ``{On the Phase Transition Towards Permanent Quark Confinement},''
\href{http://dx.doi.org/10.1016/0550-3213(78)90153-0}{{\em Nucl. Phys.} {\bf
  B138} (1978)  1--25}.

\bibitem{HO:Witten:1979ey}
E.~Witten, ``{Dyons of Charge $e\theta/2\pi$},''
\href{http://dx.doi.org/10.1016/0370-2693(79)90838-4}{{\em Phys. Lett.} {\bf
  B86} (1979)  283--287}.

\bibitem{HO:Kronheimer:1986ms}
P.~Kronheimer, ``{Monopoles and Taub-NUT metrics},'' {\em MSc. Thesis (Oxford
  University)} (1986)  .

\bibitem{HO:Biswas:1997aa}
I.~Biswas, ``{Parabolic bundles as orbifold bundles},''
  \href{http://dx.doi.org/10.1215/S0012-7094-97-08812-8}{{\em Duke Math. J.}
  {\bf 88} (1997)  305--325}.

\end{thebibliography}\endgroup

\fi
}

\newcommand{\documentfinishBBL}{
\addcontentsline{toc}{section}{References}
\ifx\ifLONG\undefined
\input{\IDvalue.separate.bbl}
\end{document}
\else
\input{\DIRvalue/\IDvalue.volume.bbl}
\fi
}

\ifx\ifLONG\undefined
\def\volcite#1{Contribution \cite{Contribution#1}}
\else 
\def\volcite#1{Chapter \ref{Chapter#1}}
\fi

\usepackage{amssymb,amsfonts,graphicx}
\usepackage{amsmath}
\usepackage{hyperref}
\usepackage{cite}
\usepackage{fullpage}

\newcommand{\HOsusy}{{\bf Q}}
\newcommand{\HObrs}{{\bf Q}_\text{B}}

\numberwithin{equation}{section}

\begin{document}
\thispagestyle{empty}
\documentheader
\else \chapterheader \fi

\section{Introduction}\label{HO:sec:int}

Four-dimensional ${\cal N}=2$ supersymemtric gauge theories are known to
be mathematically highly constrained, and yet they can
accommodate a variety of interesting physical phenomena. One can
therefore ask general questions about physics of strong gauge
interactions in these theories and expect a rather precise answer. The
first non-trivial result was obtained by Seiberg and Witten
\cite{HO:Seiberg:1994rs,HO:Seiberg:1994aj} for the structure of Coulomb branch
moduli space as well as the mass of BPS particles. By combining the
constraints from ${\cal N}=2$ supersymmetry together with
electro-magnetic duality, they determined the exact prepotential
which encodes the full low-energy effective Lagrangian, including the
contribution of instantons which were otherwise very difficult to
evaluate at that time.

Another powerful approach to 4D ${\cal N}=2$ theories is localization,
which makes use of supersymmetry to reduce the difficult problem of
infinite-dimensional path integral to a much simpler problem.
There is a class of 4D topological field theories, called topologically
twisted theories \cite{HO:Witten:1988ze}, which are obtained from
${\cal N}=2$ theories by changing the spin of fields according to their
quantum numbers under the internal symmetry $SU(2)_\text{R}$. Once the
SUSY localization is applied
to those theories, path integral can be shown to reduce to a
finite-dimensional integral on instanton moduli spaces. Nekrasov later
proposed the so-called Omega-deformation
\cite{HO:Nekrasov:2002qd,HO:Losev:1997tp,HO:Moore:1997dj,HO:Lossev:1997bz} of the
topologically twisted theories, which further simplifies the integrals on moduli
spaces by using the rotational symmetry of $\mathbb R^4$.
The resulting path integral is called Nekrasov's instanton partition
function, and is expressed as a sum over point-like instanton
configurations localized at the origin. Nekrasov's partition function
was shown to reproduce the prepotential of ${\cal N}=2$
theories in the limit of small Omega-deformation. Moreover, it has given
us a new insight into the connection between ${\cal N}=2$ gauge theories
and other branches of physics and mathematics, such as topological
strings or integrable systems.

\paragraph{Pestun's pioneering work}

Application of localization principle to quantum field theories has been
long restricted to topological field theories with scalar supersymmetry.
A major breakthrough was made by Pestun \cite{HO:Pestun:2007rz} who
constructed ${\cal N}=2$ supersymmetric gauge theories on the four-sphere
$S^4$ and derived closed formulae for partition function as well as
expectation values of certain Wilson loops \cite{HO:Pestun:2007rz}.
This article reviews his result and some of the subsequent work
on exact supersymmetric observables in ${\cal N}=2$
supersymmetric gauge theories on $S^4$ and its deformations.

The original motivation of the work \cite{HO:Pestun:2007rz} was to prove a
conjecture which arose from the study of AdS/CFT
correspondence, that the expectation values of supersymmetric circular
Wilson loops in ${\cal N}=4$ super Yang-Mills theory are given by
Gaussian matrix integral
\cite{HO:Erickson:2000af,HO:Drukker:2000rr}. Instead of topological
field theories with scalar SUSY, Pestun constructed
physical ${\cal N}=2$ SUSY theories on $S^4$ via conformal map from flat
$\mathbb R^4$. By a successful application of SUSY
localization principle, the path integral was shown to reduce to a
finite-dimensional integral. A one-parameter (mass) deformation of the
${\cal N}=4$ SYM called ${\cal N}=2^*$ theory was studied in detail, and
it was found that the integrand simplifies dramatically at a special
value of the mass. In this way, it was analytically shown that the $S^4$
partition function is precisely given by a Gaussian matrix integral
\cite{HO:Pestun:2007rz,HO:Okuda:2010ke}. See \volcite{ZA} for more
detail on the application of localization to the problems in AdS/CFT.

Pestun's work is the first nontrivial example in which a coherent and
fully explicit prescription was given for physical supersymmetric models
on curved spaces, from the construction of theories to the evaluation of
supersymmetric observables. Exact formulae were
obtained later for partition functions of supersymmetric gauge theories on
$S^3$ \cite{HO:Kapustin:2009kz,HO:Jafferis:2010un,HO:Hama:2010av},
$S^2$ \cite{HO:Benini:2012ui,HO:Doroud:2012xw} and $S^5$
\cite{HO:Kallen:2012va,HO:Kim:2012ava,HO:Kim:2012qf} by following basically the
same program. Together with the supersymmetric partition functions on
$S^1\times S^d$ called superconformal indices, the sphere partition
functions are now regarded as powerful analytic tools to
explore non-perturbative aspects of SUSY gauge theories.
In particular, for CFTs with right number of supersymmetry in even
dimensions, it was shown that the sphere partition function is protected from
regularization ambiguity and computes the K\"ahler potential for the space
of marginal couplings
\cite{HO:Gerchkovitz:2014gta,HO:Gomis:2014woa}.

Important applications of Pestun's result have been made for a
family of 4D ${\cal N}=2$ theories of ``class S'' \cite{HO:Gaiotto:2009we},
that are known to show up on the worldvolume of multiple M5-branes
($5+1$-dimensional object in M-theory) wrapped on punctured Riemann
surfaces. In particular, Alday, Gaiotto and Tachikawa (AGT) discovered a
surprising correspondence between exact $S^4$ partition functions of the
class S superconformal theories for two M5-branes and correlation
functions of 2D Liouville conformal field theory \cite{HO:Alday:2009aq}
(see \volcite{TA}).
Generalization to gauge groups of higher rank and Toda conformal
field theories was soon proposed by Wyllard \cite{HO:Wyllard:2009hg}. This
discovery brought us with another new insight
into the mathematical structure underlying 4D ${\cal N}=2$ gauge
theories. It also triggered an extensive study of similar correspondences
between quantum field theories in different dimensions that follow from
compactifications of multiple M5-branes.

\paragraph{Squashing}

Supersymmetric gauge theories and exact physical observables have also
been studied on manifolds which are less symmetric than sphere. One
motivation for this generalization arose from the AGT relation, since
the partition function on the round $S^4$ was shown to correspond
to Toda CFTs at a special (self-dual) value of the coupling, $b=1$.
Nontrivial results along this line of generalization were first obtained
in \cite{HO:Hama:2011ea} and \cite{HO:Imamura:2011wg} for 3D ${\cal N}=2$
supersymmetric theories on certain squashed spheres with a background vector
field turned on. The supersymemtry there is characterized by generalized
Killing spinors with a specific coupling to the vecor field.

For theories with different amount of SUSY and in other dimensions, the
most natural framework to explore supersymmetric curved
backgrounds is off-shell supergravity \cite{HO:Festuccia:2011ws}, See
\volcite{DU}. For 4D
${\cal N}=2$ theories this idea was employed in \cite{HO:Hama:2012bg} to
construct supersymmetric ellipsoid backgrounds, which depend on a squashing
parameter $b$ measuring the deformation from the round sphere geometry. The
partition function on this background was shown to reproduce the
correlators of Toda CFTs at general values of the coupling.
The rigid supersymmetric backgrounds were systematically classified and
deformations of $S^4$ were studied within ${\cal N}=1$ off-shell
supergravity in
\cite{HO:Samtleben:2012gy,HO:Dumitrescu:2012ha,HO:Cassani:2012ri,HO:Liu:2012bi,HO:Dumitrescu:2012at},
and in \cite{HO:Klare:2013dka} within ${\cal N}=2$ supergravity.
Different versions of deformations of the round $S^4$ have been studied in
\cite{HO:Nosaka:2013cpa,HO:Cabo-Bizet:2014nia}, while the backgrounds of
other topologies, such as products of spheres and $AdS$ spaces,
have been studied in
\cite{HO:Bawane:2014uka,HO:Sinamuli:2014lma,HO:Murthy:2015yfa,HO:Gupta:2015gga,HO:Butter:2015tra},
where the results have been used to study the
loop correction to the entropy of certain charged black holes.
Supersymmetric deformations of the round sphere geometry have
also been applied to the computation of R\'enyi entropy in gauge
theories in $D=3,4,5$; see
\cite{HO:Nishioka:2013haa,HO:Huang:2014pda,HO:Crossley:2014oea,HO:Hama:2014iea,HO:Zhou:2015cpa}.

\paragraph{Supersymmetric observables}

Localization techniques have also been applied to compute expectation
values of various supersymmetric observables. An important class of
observables in 4D ${\cal N}=2$ theories are supersymmeric Wilson and 't
Hooft loop operators, defined from the worldlines of electrically or
magnetically charged particles. It is a remarkable feature of ${\cal N}=2$
supersymmetric theories that one can make quantitative statements about
properties of these particles, in particular how they are exchanged
among each other under S-duality \cite{HO:Gaiotto:2009we}. Also, a number
of nontrivial conjectures on the expectation values of loop operators
have been proposed from AGT relation and checked explicitly
\cite{HO:Alday:2009fs,HO:Drukker:2009id,HO:Gomis:2011pf}. The effect of
deformations of the theories on Wilson loop and local observables were
studied in \cite{HO:Fucito:2015ofa}.

Another important class of nonlocal operators are surface operators,
which have two dimensional worldvolume inside four dimensions. See
\cite{HO:Gukov:2014gja} for a review. They are defined either by introducing
two-dimensional field theory degrees of freedom on the surface or by
imposing singular behavior on gauge and other fields along the
surface. They were first introduced in \cite{HO:Gukov:2006jk} in the
study of geometric Langlands program within the framework of 4D
${\cal N}=4$ SYM theory. Interesting progress has been made for
surface operators in ${\cal N}=2$ supersymmetric theories through the
comparison of the gauge theory analysis with the results from
topological string or predictions from AGT relation \cite{HO:Alday:2009fs,HO:Alday:2010vg,HO:Dimofte:2010tz,HO:Awata:2010bz,HO:Kozcaz:2010yp,HO:Wyllard:2010rp,HO:Wyllard:2010vi,HO:Kanno:2011fw,HO:Gomis:2014eya,HO:Nawata:2014nca}.

\paragraph{Conventions}

Throughout this article, we use the indices $\alpha,\beta,\cdots$ and
$\dot\alpha,\dot\beta,\cdots$ for 4D chiral and anti-chiral spinors.
The indices are raised and lowered by the antisymmetric invariant
tensors $\epsilon^{\alpha\beta}$, $\epsilon^{\dot\alpha\dot\beta}$,
$\epsilon_{\alpha\beta}$, $\epsilon_{\dot\alpha\dot\beta}$ with nonzero
elements
\begin{equation}
 \epsilon^{12}=-\epsilon^{21}=-\epsilon_{12}=\epsilon_{21}=1.
\end{equation}
Following Wess-Bagger \cite{HO:Wess:1992cp} we suppress the pairs of
undotted indices contracted in the up-left, down-right order, or
pairs of dotted indices contracted in the down-left, up-right order.
We also use the set of $2\times2$ matrices
$(\sigma^a)_{\alpha\dot\alpha}$ and $(\bar\sigma^a)^{\dot\alpha\alpha}$
with $a=1,\cdots,4$ satisfying standard algebras. In terms of Pauli's
matrices $\boldsymbol\tau^a$ they are given by
\begin{equation}
\begin{array}{lll}
 \sigma^a = -i\boldsymbol\tau^a, &
 \bar\sigma^a = i\boldsymbol\tau^a, &
 (a=1,2,3)\\
 \sigma^4 = 1, &
 \bar\sigma^4 = 1. &
\end{array}
\end{equation}
We also use
$\sigma_{ab}=\frac12(\sigma_a\bar\sigma_b-\sigma_b\bar\sigma_a)$ and
$\bar\sigma_{ab}=\frac12(\bar\sigma_a\sigma_b-\bar\sigma_b\sigma_a)$.
Note that $\sigma_{ab}$ is anti-self-dual,
i.e. $\sigma_{ab}=-\frac12\varepsilon_{abcd}\sigma_{cd}$, while
$\bar\sigma_{ab}$ is self-dual.

For 4D ${\cal N}=2$ theories on flat space, supersymmetry is
parametrized by constant spinors $\xi_{\alpha A}$ and $\bar\xi^{\dot\alpha}_A$.
The index $A=1,2$ indicates that they transform as doublet under $SU(2)$
R-symmetry which commutes with the generators of Poincar\'e symmetry but
rotates the supercharges. In addition, $\xi_{\alpha A}$ and
$\bar\xi^{\dot\alpha}_A$ carry $U(1)$ R-charges $+1$ and $-1$.
Throughout this article, these SUSY parameters are Grassmann-even quantities.

\section{Construction of theories}\label{HO:sec:cns}

Here we review the construction of ${\cal N}=2$ supersymmetric gauge
theories on $S^4$ using off-shell supergravity. We then present a number
of nontrivial supergravity backgrounds with rigid supersymmetry,
including the supersymmetric deformation of $S^4$ into ellipsoids.

\subsection{Conformal Killing spinors on $S^4$}\label{HO:sec:cks}

As the round $S^4$ is conformally flat, 4D ${\cal N}=2$ superconformal
theories can be constructed on $S^4$ by a conformal map from flat $\mathbb R^4$.
Let $\ell$ be the radius of $S^4$. The superconformal symmetry is then
described by conformal Killing spinors satisfying
\begin{align}
 D_m\xi_A &\equiv
 \left(\partial_m+\frac14\Omega_m^{ab}\sigma_{ab}\right)\xi_A
 = -i\sigma_m\bar\xi'_A,&
 D_m\bar\xi'_A &= -\frac{i}{4\ell^2}\bar\sigma_m\xi_A,
 \nonumber\\
 D_m\bar\xi_A &\equiv
 \left(\partial_m+\frac14\Omega_m^{ab}\bar\sigma_{ab}\right)\bar\xi_A
 = -i\bar\sigma_m\xi'_A,&
 D_m\xi'_A &= -\frac{i}{4\ell^2}\sigma_m\bar\xi_A.
\label{HO:eq:cks}
\end{align}
This is a coupled first-order differential equation for 16 spinor
components, and therefore has 16 independent solutions corresponding to
the fermionic generators of the 4D ${\cal N}=2$ superconformal algebra.
Lagrangian theories of vector multiplets and massless hypermultiplets
are all superconformal at the classical level, so they can be unambiguously
defined on the round $S^4$ in this way. For massive theories on
$S^4$, the superconformal symmetry is broken to a subgroup $OSp(2|4)$.
This means that the mass terms are constructed in such a way that a
subset of supercharges corresponding to the Killing spinors
\begin{equation}
 D_m\xi_A=-\frac i{2\ell}\sigma_m\bar\xi_B\cdot t^B_{~A},\quad
 D_m\bar\xi_A=-\frac i{2\ell}\bar\sigma_m\xi_B\cdot\bar t^B_{~A}
\label{HO:eq:ksm}
\end{equation}
is preserved. Here $t,\bar t$ are constant traceless $U(2)$ matrices
satisfying $t\bar t=\bar tt={\bf 1}$. They can be brought into a
standard form, say $t=\bar t=\boldsymbol{\tau}_3$, using R-symmetry.

\subsection{Generalized Killing spinors and ${\cal N}=2$ Supergravity}\label{HO:sec:gks}

\begin{table}[t]
\begin{center}
\begin{tabular}{|cl|c|}
\hline
\multicolumn{2}{|c|}{bosons} & $q_\text{R}$ \\
\hline
$g_{mn}$ & metric & $0$ \\
$(V_m)^A_{~B}$ & gauge field for $SU(2)_\text{R}$ & $0$ \\
$\tilde V_m$ & gauge field for $U(1)_\text{R}$ & $0$ \\
$T_{mn}$ & anti-self-dual tensor & $+2$ \\
$\bar T_{mn}$ & self-dual tensor & $-2$\\
$\tilde M$ & scalar & $0$ \\
\hline
\end{tabular}
\begin{tabular}{|cl|c|}
\hline
\multicolumn{2}{|c|}{fermions} & $q_\text{R}$ \\
\hline
$\psi_{mA}$ & chiral gravitino & $+1$ \\
$\bar\psi_{mA}$ & anti-chiral gravitino & $-1$ \\
$\eta_A$ & chiral spinor & $+1$ \\
$\bar\eta_A$ & anti-chiral spinor & $-1$ \\
\hline
\multicolumn{3}{c}{}\\
\multicolumn{3}{c}{}\\
\end{tabular}
\caption{fields and their $U(1)_\text{R}$ charges $q_\text{R}$
in off-shell 4D ${\cal N}=2$ supergravity\label{HO:tbl:sgr}}
\end{center}
\end{table}
Off-shell supergravity allows to construct supersymmetric field theories
on more general curved backgrounds \cite{HO:Festuccia:2011ws}. The
independent fields in the standard gravity multiplet (also called {\it
Weyl multiplet}) in 4D ${\cal N}=2$ supergravity
\cite{HO:deWit:1979dzm,HO:deWit:1980lyi,HO:deWit:1983xe,HO:deWit:1984rvr}
(see also reviews \cite{HO:Mohaupt:2000mj,HO:Freedman:2012zz}) are listed
in Table \ref{HO:tbl:sgr}. Supergravity backgrounds are specified by
the classical values of all the bosonic fields, while the fermionic
fields are all taken to vanish. A background is supersymmetric if the
local SUSY variation of fermions, (we quote the formula from
\cite{HO:Freedman:2012zz} with certain rescalings of fields)
\begin{align}
 \HOsusy\psi_{mA} &\,=\,
 D_m\xi_A+T^{kl}\sigma_{kl}\sigma_m\bar\xi_A
 +i\sigma_m\bar\xi'_A,
 \nonumber \\
 \HOsusy\bar\psi_{mA} &\,=\,
 D_m\bar\xi_A+\bar T^{kl}\bar\sigma_{kl}\bar\sigma_m\xi_A
 +i\bar\sigma_m\xi'_A,
 \nonumber \\
 \HOsusy\eta_A &\,=\,
 8\sigma^{mn}\sigma^l\bar\xi_AD_lT_{mn}
 +16iT^{kl}\sigma_{kl}\xi'_A
 -3\tilde M\xi_A
 +2i\sigma^{mn}\xi_B(V_{mn})^B_{~A}
 +4i\sigma^{mn}\xi_A\tilde V_{mn},
 \nonumber \\
 \HOsusy\bar\eta_A &\,=\,
 8\bar\sigma^{mn}\bar\sigma^l\xi_AD_l\bar T_{mn}
 +16i\bar T^{kl}\bar\sigma_{kl}\bar\xi'_A
 -3\tilde M\bar\xi_A
 +2i\bar\sigma^{mn}\bar\xi_B(V_{mn})^B_{~A}
 -4i\bar\sigma^{mn}\bar\xi_A\tilde V_{mn},
\label{HO:eq:ksg}
\end{align}
all vanish for a suitable choice of spinor fields $(\xi_A,\bar\xi_A)$ and
$(\xi'_A,\bar\xi'_A)$. Here the covariant derivatives are with respect
to the local Lorentz as well as $SU(2)\times U(1)$ R-symmetries. For example,
\begin{equation}
 D_m\xi_A\equiv \left(\partial_m+\frac14\Omega_m^{ab}\sigma_{ab}\right)\xi_A
 +i\xi_B(V_m)^B_{~A}-i\tilde V_m\xi_A.
\end{equation}
We also denoted the $U(1)_\text{R}$ gauge field strength by
$\partial_m\tilde V_n-\partial_n\tilde V_m\equiv \tilde V_{mn}$ and
similarly for the $SU(2)_\text{R}$ field strength
$(V_{mn})^B_{~A}$. With the simplifying assumption
\begin{equation}
 \tilde V_m=0,
\end{equation}
the above BPS condition can be transformed into the form presented in
\cite{HO:Hama:2012bg},
\begin{align}
 D_m\xi_A+T^{kl}\sigma_{kl}\sigma_m\bar\xi_A
 &\,=\, -i\sigma_m\bar\xi'_A,
 \nonumber \\
 D_m\bar\xi_A+\bar T^{kl}\bar\sigma_{kl}\bar\sigma_m\xi_A
 &\,=\, -i\bar\sigma_m\xi'_A,
 \nonumber \\
 \sigma^m\bar\sigma^nD_mD_n\xi_A+4D_lT_{mn}\sigma^{mn}\sigma^l\bar\xi_A
 &\,=\, M\xi_A,
 \nonumber \\
 \bar\sigma^m\sigma^nD_mD_n\bar\xi_A
 +4D_l\bar T_{mn}\bar\sigma^{mn}\bar\sigma^l\xi_A
 &\,=\, M\bar\xi_A,
\label{HO:eq:ksh}
\end{align}
where $M\equiv \tilde M-\frac13R$.
This gives a consistent generalization of the conformal Killing spinor
equation (\ref{HO:eq:cks}) on $S^4$. Hereafter we use $M$ rather than
$\tilde M$ in accordance with \cite{HO:Hama:2012bg}, but note that the
latter has a better transformation property under Weyl rescaling. The
equations (\ref{HO:eq:ksh}) are invariant under
$g_{mn}\to e^{2\rho}g_{mn}$ if accompanied by
\begin{alignat}{4}
 \xi_A&\to e^{\frac12\rho}\xi_A,& \quad
 \xi'_A&\to e^{-\frac12\rho}\xi'_A,& \quad
 T_{mn}&\to e^{-\rho}T_{mn},& \quad
 \tilde M&\to e^{-2\rho}\tilde M,
\nonumber \\
 \bar\xi_A&\to e^{\frac12\rho}\bar\xi_A,& \quad
 \bar\xi'_A&\to e^{-\frac12\rho}\bar\xi'_A,& \quad
 \bar T_{mn}&\to e^{-\rho}\bar T_{mn}.
\end{alignat}

\subsection{Transformation laws and Lagrangians}\label{HO:sec:lag}

Supergravity also gives a description of local SUSY-invariant couplings
of matter systems to gravity. By sending the Newton constant to
zero in such a description, one can decouple gravity from the matter and
treat the fields in gravity multiplet as classical background fields. In
this way one can systematically construct rigid SUSY theories on various
curved backgrounds.

Vector multiplet consists of a gauge field $A_m$, scalars $\phi,\bar\phi$,
gauginos $\lambda_{\alpha A},\bar\lambda_{\dot\alpha A}$ and an
$SU(2)_\text{R}$-triplet auxiliary scalar $D_{AB}$. They transform under
supersymmetry as
\begin{align}
 \HOsusy A_m &\,=\,
 i\xi^A\sigma_m\bar\lambda_A-i\bar\xi^A\bar\sigma_m\lambda_A,
 \nonumber \\
 \HOsusy\phi &\,=\, -i\xi^A\lambda_A,
 \nonumber \\
 \HOsusy\bar\phi &\,=\, +i\bar\xi^A\bar\lambda_A,
 \nonumber \\
 \HOsusy\lambda_A &\,=\,
 \tfrac12\sigma^{mn}\xi_A(F_{mn}+8\bar\phi T_{mn})
 +2\sigma^m\bar\xi_AD_m\phi+\sigma^mD_m\bar\xi_A\phi
 +2i\xi_A[\phi,\bar\phi]+D_{AB}\xi^B,
 \nonumber \\
 \HOsusy\bar\lambda_A &\,=\,
 \tfrac12\bar\sigma^{mn}\bar\xi_A(F_{mn}+8\phi\bar T_{mn})
 +2\bar\sigma^m\xi_AD_m\bar\phi+\bar\sigma^mD_m\xi_A\bar\phi
 -2i\bar\xi_A[\phi,\bar\phi]+D_{AB}\bar\xi^B,
 \nonumber \\
 \HOsusy D_{AB} &\,=\,
 -i\bar\xi_A\bar\sigma^mD_m\lambda_B
 -i\bar\xi_B\bar\sigma^mD_m\lambda_A
 +i\xi_A\sigma^mD_m\bar\lambda_B
 +i\xi_B\sigma^mD_m\bar\lambda_A
 \nonumber \\ & \hskip5.5mm
 -2[\phi,\bar\xi_A\bar\lambda_B+\bar\xi_B\bar\lambda_A]
 +2[\bar\phi,\xi_A\lambda_B+\xi_B\lambda_A].
\label{HO:eq:qvm}
\end{align}
Note that the following combination of vector and scalar fields is
$\HOsusy$-invariant,
\begin{equation}
 \hat\Phi~\equiv~ 2i\xi^A\xi_A\bar\phi-2i\bar\xi^A\bar\xi_A\phi
 -2i\bar\xi^A\bar\sigma^m\xi_A A_m,
\label{HO:eq:htp}
\end{equation}
which will become important later.
SUSY invariant Yang-Mills kinetic Lagrangian reads
\begin{align}
 {\cal L}_\text{YM} &\,=\,
 \frac1{g^2}\text{Tr}\Big(\tfrac12F_{mn}F^{mn}
 +16F_{mn}(\bar\phi T^{mn}+\phi\bar T^{mn})
 +64\bar\phi^2T_{mn}T^{mn}+64\phi^2\bar T_{mn}\bar T^{mn}
 \nonumber \\ & \hskip13mm
 -4D_m\bar\phi D^m\phi+2M\bar\phi\phi
 -2i\lambda^A\sigma^mD_m\bar\lambda_A
 -2\lambda^A[\bar\phi,\lambda_A]+2\bar\lambda^A[\phi,\bar\lambda_A]
 \nonumber \\ & \hskip13mm
 +4[\phi,\bar\phi]^2-\tfrac12D^{AB}D_{AB}
 \Big)
 +\frac{i\theta}{32\pi^2}\text{Tr}\Big(\varepsilon^{klmn}F_{kl}F_{mn}\Big)\,.
\label{HO:eq:lym}
\end{align}
One instanton factor is $q=e^{2\pi i\tau}$ with
$\tau=\frac{\theta}{2\pi}+\frac{4\pi i}{g^2}$.

For $U(1)$ vector multiplets one can also construct a Feyet-Illiopoulos type
invariant. Let $w^{AB}=w^{BA}$ be an $SU(2)_\text{R}$-triplet background
field satisfying
\begin{align}
 w^{AB}\xi_B &\,=\,
 \frac12\sigma^nD_n\bar\xi^A+2T_{kl}\sigma^{kl}\xi^A,
 \nonumber \\
 w^{AB}\bar\xi_B &,=\,
 \frac12\bar\sigma^nD_n\xi^A+2\bar T_{kl}\bar\sigma^{kl}\bar\xi^A.
\label{HO:eq:wab}
\end{align}
Then the following is SUSY-invariant.
\begin{equation}
 {\cal L}_\text{FI} =
 \zeta\left\{
 w^{AB}D_{AB}-M(\phi+\bar\phi)-64\phi T^{kl}T_{kl}
 -64\bar\phi\bar T^{kl}\bar T_{kl}-8F^{kl}(T_{kl}+\bar T_{kl})
 \right\}\,.
\label{HO:eq:lfi}
\end{equation}
Note that this term breaks the conformal invariance. By comparing with
the Killing spinor equation (\ref{HO:eq:ksm}), one finds
$t_{AB}=\bar t_{AB}=i\ell w_{AB}$  on the round $S^4$ of radius $\ell$,.

The system of $r$ hypermultiplets consists of scalars $q_{IA}$ and
fermions $\psi_{\alpha I},\bar\psi_{\dot\alpha I}$, with
$I=1,\cdots,2r$. The scalars obey the reality condition
\begin{equation}
 (q_{IA})^\dagger~=~ q^{AI}~=~\epsilon^{AB}\Omega^{IJ}q_{JB},
\label{HO:eq:qrl}
\end{equation}
where $\Omega^{IJ}$ is the real antisymmetric $Sp(r)$-invariant tensor
satisfying
\[
 (\Omega^{IJ})^\ast=-\Omega_{IJ},~\Omega^{IJ}\Omega_{JK}=\delta^I_K.
\]
The tensor $\Omega^{IJ}$ and its inverse are used to raise or lower the
$Sp(r)$ indices. The pair of $Sp(r)$ indices will be suppressed in the
following when
contracted in the top-left, bottom-right order, like $q^{AI}q_{IA}=q^Aq_A$.
The hypermultiplet fields can be coupled to vector multiplet by
embedding the gauge group into $Sp(r)$. The covariant derivative of
$q_{IA}$, for example, is then given by
\begin{equation}
 D_mq_{IA}\equiv \partial_mq_{IA}-i(A_m)_I^{~J}q_{JA}
 +iq_{IB}(V_m)^B_{~A}.
\end{equation}

It is a little intricate to write down an off-shell SUSY transformation
rule for hypermultiplet fields explicitly. As is well known, for rigid
${\cal N}=2$ SUSY theories with hypermultiplets on flat space, there is
no formalism which realizes all the 8 supercharges at once with finite
number of auxiliary fields. However, when applying localization method,
one always picks up one of the supercharges corresponding to a particular
choice of Killing spinor $\xi_A,\bar \xi_A$, and requires that particular
supercharge to be realized off-shell. What we will present here is an
off-shell realization of just one supercharge.

To balance the number of bosons and fermions in hypermultiplet, we need
to introduce the auxiliary scalar fields $F_{I\check A}$, where $I$ is
the $Sp(r)$ index and $\check A=1,2$ is a new auxiliary index. We also
introduce \cite{HO:Hama:2012bg} the spinor fields
$\check\xi_{\check A},\bar{\check\xi}_{\check A}$ satisfying
\begin{align}
 \xi_A\check\xi_{\check B}-\bar\xi_A\bar{\check\xi}_{\check B} &\,=\, 0,
 \nonumber \\
 \xi^A\xi_A+\bar{\check\xi}^{\check A}\bar{\check\xi}_{\check A} &\,=\, 0,
 \nonumber \\
 \bar\xi^A\bar\xi_A+\check\xi^{\check A}\check\xi_{\check A} &\,=\, 0,
 \nonumber \\
 \xi^A\sigma^m\bar\xi_A+\check\xi^{\check A}\sigma^m\bar{\check\xi}_A &\,=\, 0.
\label{HO:eq:xck}
\end{align}
A solution to the above conditions is given by
\begin{equation}
 \check\xi_{\check A}= c^{\frac12}\xi_A,~~~
 \bar{\check\xi}_{\check A}=-c^{-\frac12}\bar\xi_A~~(A=\check A=1,2)~~~
 \text{where}~~~
 c=-\frac{\bar\xi^A\bar\xi_A}{\xi^B\xi_B}.
\label{HO:eq:cxi}
\end{equation}
There are more solutions since the equations (\ref{HO:eq:xck}) is
invariant under local $SL(2)$ transformations acting $\check\xi_{\check A}$
and $\bar{\check\xi}_{\check A}$ through the index $\check A$, but one
can show the solution is unique up to this $SL(2)$.
Using them the SUSY transformation rule for hypermultiplet
can be expressed as follows,
\begin{align}
 \HOsusy q_A &\,=\, -i\xi_A\psi+i\bar\xi_A\bar\psi,
 \nonumber \\
 \HOsusy\psi &\,=\, 2\sigma^m\bar\xi_AD_mq^A+\sigma^mD_m\bar\xi_Aq^A
 -4i\xi_A\bar\phi q^A+2\check\xi_{\check A} F^{\check A},
 \nonumber \\
 \HOsusy\bar\psi &\,=\, 2\bar\sigma^m\xi_AD_mq^A+\bar\sigma^mD_m\xi_Aq^A
 -4i\bar\xi_A\phi q^A+2\bar{\check\xi}_{\check A}F^{\check A},
 \nonumber \\
 \HOsusy F_{\check A} &\,=\,
  i\check\xi_{\check A}\sigma^mD_m\bar\psi
 -2\check\xi_{\check A}\phi\psi
 -2\check\xi_{\check A}\lambda_Bq^B+2i\check\xi_{\check A}
  (\sigma^{kl}T_{kl})\psi
 \nonumber \\ & \hskip4.4mm
 -i\bar{\check\xi}_{\check A}\bar\sigma^mD_m\psi
 +2\bar{\check\xi}_{\check A}\bar\phi\bar\psi
 +2\bar{\check\xi}_{\check A}\bar\lambda_Bq^B
 -2i\bar{\check\xi}_{\check A}(\bar\sigma^{kl}\bar T_{kl})\bar\psi\,.
\end{align}
Similar off-shell transformation rule was used in \cite{HO:Pestun:2007rz}
for 4D ${\cal N}=4$ SYM theory using Berkovits construction of 10D
${\cal N}=1$ SYM theory \cite{HO:Berkovits:1993hx}. The SUSY invariant
kinetic Lagrangian is
\begin{align}
 {\cal L}_\text{mat} &\,=\,
 \tfrac12D_mq^AD^mq_A-q^A\{\phi,\bar\phi\}q_A+\tfrac i2q^AD_{AB}q^B
 +\tfrac18(M+R)q^Aq_A \nonumber \\& \hskip5mm
 -\tfrac i2\bar\psi\bar\sigma^mD_m\psi
 -\tfrac12\psi\phi\psi+\tfrac12\bar\psi\bar\phi\bar\psi
 +\tfrac i2\psi\sigma^{kl}T_{kl}\psi
 -\tfrac i2\bar\psi\bar\sigma^{kl}\bar T_{kl}\bar\psi
 \nonumber \\&\hskip5mm
 -q^A\lambda_A\psi+\bar\psi\bar\lambda_Aq^A
 -\tfrac12F^{\check A}F_{\check A}.
\label{HO:eq:lmt}
\end{align}
The auxiliary symmetry transforms $F_{I\check A}$, $\check\xi_{\check A}$
and $\bar\xi_{\check A}$ all as doublets, and it is actually $SU(2)$ since
we need to impose $F_{I\check A}$ a reality condition similar to
(\ref{HO:eq:qrl}). To complete the off-shell formalism for
hypermultiplets, one needs to specify the background gauge field
${(\check V_m)^{\check A}}_{\check B}$ for this auxiliary
symmetry which we call $SU(2)_{\check{\text R}}$.

The commutant of the gauge group within $Sp(r)$ gives the global
symmetry. One can introduce the mass for hypermultiplets by coupling an
abelian subgroup of the global symmetry to background vector multiplets.
Mass parameters are identified with the constant value of their scalar
components $\phi,\bar\phi$. They have to be chosen not to break
supersymmetry, so the fermion components of the background vector
multiplet must have vanishing SUSY variation. The classical values
\begin{equation}
 \phi=\bar\phi=\text{constant},\quad
 D_{AB}=2w_{AB}\phi
\end{equation}
preserve the supersymmetry if the corresponding Killing spinor satisfy
(\ref{HO:eq:wab}).

The square of supersymmetry $\HOsusy$ yields a sum of bosonic symmetry
transformations including the translation by
$v^m \equiv 2\bar\xi^A\bar\sigma^m\xi_A$,
\begin{align}
 \HOsusy^2 \,=\,&\,
 i{\cal L}_v
 \nonumber \\ &
 +\text{Gauge}\big[2\phi\bar\xi^A\bar\xi_A-2\bar\phi\xi^A\xi_A+v^mA_m\big]
 \nonumber \\ &
 +\text{Lorentz}\big[D_{[a}v_{b]}+v^m\Omega_{mab}\big]
 \nonumber \\ &
 +\text{Scale}\big[-\tfrac i2\xi^A\sigma^mD_m\bar\xi_A
                   -\tfrac i2D_m\xi^A\sigma^m\bar\xi_A \big]
 \nonumber \\ &
 +\text{R}_{U(1)}
 \big[-\tfrac i4\xi^A\sigma^mD_m\bar\xi_A
      +\tfrac i4D_m\xi^A\sigma^m\bar\xi_A \big]
 \nonumber \\ & +\text{R}_{SU(2)}
 \big[-i\xi_{(A}\sigma^mD_m\bar\xi_{B)}
      +iD_m\xi_{(A}\sigma^m\bar\xi_{B)}+v^mV_{mAB}\big]
 \nonumber \\ &
 +\check{\text{R}}_{SU(2)}
 \big[2i\check\xi_{(A}\sigma^mD_m\bar{\check\xi}_{B)}
     -2iD_m\check\xi_{(A}\sigma^m\bar{\check\xi}_{B)}
 \nonumber \\ & \hskip14mm
     +4i\check\xi_{(A}\sigma^{kl}T_{kl}\check\xi_{B)}
     -4i\bar{\check\xi}_{(A}\bar\sigma^{kl}\bar T_{kl}\bar{\check\xi}_{B)}
     +v^m\check V_{mAB}
 \big]\,.
\end{align}
Note that the Killing spinor $(\xi_A,\bar\xi_A)$, the auxiliary spinor
$(\check\xi_{\check A},\bar{\check\xi}_{\check A})$ as well as all the
background fields belonging to the gravity multiplet have to be invariant
under $\HOsusy^2$. This can be used to determine the form of
${(\check V_m)^{\check A}}_{\check B}$. Note also that, if one wants to
introduce the mass or FI terms into the theory, the Killing spinor has
to satisfy an extra condition (\ref{HO:eq:wab}). This implies
\begin{equation}
 \xi^A\sigma^mD_m\bar\xi_A~=~\bar\xi^A\bar\sigma^mD_m\xi_A~=~0,
\end{equation}
so that $\HOsusy^2$ does not yield scale or $U(1)_\text{R}$
transformations.

\subsection{Examples of SUSY backgrounds}\label{HO:sec:sbg}

Let us review here some important examples of classical supergravity
backgrounds with rigid SUSY.

\paragraph{Topological twist}

It is known that 4D ${\cal N}=2$ theories can be put on any 4D spaces
preserving a single scalar supercharge by a procedure called
Donaldson-Witten topological twist \cite{HO:Witten:1988ze}. In the
supergravity framework, topological twist corresponds to turning on a
background $SU(2)_\text{R}$ gauge field which equals the self-dual part
of spin connection,
\begin{equation}
  \frac14\Omega_m^{ab}(\bar\sigma_{ab})^A_{~B}+i(V_m)^A_{~B}=0,
\end{equation}
so that the constant spinor
$\xi_{\alpha A}=0,~\bar\xi^{\dot\alpha}_A=\delta^{\dot\alpha}_A$ 
satisfies the Killing spinor equation (\ref{HO:eq:ksg}).
The supersymmetry $\HOsusy$ is nilpotent up to gauge transformations, so
that one can define physical observables by cohomology of $\HOsusy$ acting
on gauge-invariant operators.

The choice of the background $SU(2)_\text{R}$ gauge field allows one to
identify the indices $A,B,\cdots$ with the dotted spinor indices. The
chiral gaugino $\lambda_{\alpha A}$ then turns into a vector $\psi_m$
which is the superpartner of $A_m$ under $\HOsusy$, whereas the
anti-chiral gaugino $\bar\lambda^{\dot\alpha}_A$ gives rise to a scalar
$\eta$ and a self-dual tensor $\chi_{mn}^+$. The fermion $\chi_{mn}^+$
and its superpartner play the role of Lagrange multiplier which reduces the
path integral over the gauge field to a finite-dimensional moduli space
of instanton configurations satisfying
$\frac12\varepsilon_{klmn}F^{mn}=-F_{kl}$.
The contribution from $k$-instanton configurations is weighted by
$e^{2\pi i k\tau}=q^k$ since the SYM action can be written as
\begin{equation}
 S_\text{YM} = 2\pi i\tau\cdot
 \frac1{8\pi^2}\int\text{Tr}F\wedge F
 +\HOsusy(\cdots)\,.
\end{equation}
Similarly, by setting $SU(2)_\text{R}$ gauge field equal to the
anti-self-dual part of spin connection, one obtains a supersymmetric
background corresponding to anti-twisted theory for which the path
integrals localize to moduli space of anti-instantons.

\paragraph{Omega backgrounds}

Omega background is a deformation of topologically twisted theory such
that $\HOsusy$ is not nilpotent but squares to an isometry of the
background metric. The simplest example is the Omega-deformation of
flat space often denoted as $\mathbb R^4_{\epsilon_1,\epsilon_2}$.
Path integrals of gauge theories on such a background reduce to
equivariant integrals on instanton moduli space, that is the problem of
counting the configurations of point-like instantons localized at the
origin, and gives the definition of Nekrasov's instanton partition
function \cite{HO:Nekrasov:2002qd,HO:Losev:1997tp,HO:Moore:1997dj,HO:Lossev:1997bz}.

To be a little more explicit, the Omega background
$\mathbb R^4_{\epsilon_1,\epsilon_2}$ is characterized by a scalar
supercharge which squares to a rotation,
\begin{equation}
 \HOsusy^2=i{\cal L}_v+(\cdots),\quad
 v \equiv
 \epsilon_1\left(
  x_1\frac\partial{\partial x_2}-x_2\frac\partial{\partial x_1}
 \right)
+ \epsilon_2\left(
  x_3\frac\partial{\partial x_4}-x_4\frac\partial{\partial x_3}
 \right)\,.
\end{equation}
To realize it within the supergravity framework, one chooses the Killing
spinor with constant $\bar\xi_A$ as before, and also a nonvanishing $\xi_A$ so
that $2\bar\xi^A\bar\sigma^m\xi_A=v^m$ holds. More explicitly,
\begin{equation}
 \bar\xi^{\dot\alpha}_A = \frac1{\sqrt2}\delta^{\dot\alpha}_A,
 \quad
 \xi_{\alpha A} =
 -\frac12v_m(\sigma^m)_{\alpha\dot\alpha}\bar\xi^{\dot\alpha}_A\,.
\label{HO:eq:kso}
\end{equation}
In order for this to satisfy the equation (\ref{HO:eq:ksh}) one needs to
put
\begin{equation}
 M=\bar T_{kl}=0,~~~T_{kl}=-\frac18{D_{[k}v_{l]}}^-~~\left(\text{or}~
 \frac12T_{kl}{\rm d}x^k{\rm d}x^l
 =\frac{\epsilon_2-\epsilon_1}{16}
 ({\rm d}x^1{\rm d}x^2-{\rm d}x^3{\rm d}x^4)\right)
\label{HO:eq:bgo}
\end{equation}
Note that for $\epsilon_1=\epsilon_2$ no background auxiliary fields
need to be turned on. A related remark is that the orientation reversal
of one of the coordinate axes (``parity'') leads to the sign flip of either
$\epsilon_1$ or $\epsilon_2$, but at the same time flips the definition
of chirality for spinors. Therefore, twisted theory on
$\mathbb R^4_{\epsilon_1,\epsilon_2}$ and anti-twisted theory on
$\mathbb R^4_{\epsilon_1,-\epsilon_2}$ are related by parity.

For the choice of Killing spinor (\ref{HO:eq:kso}), the simplest
solution to the equation (\ref{HO:eq:xck}) is
\begin{equation}
 \check\xi^{\check A}_\alpha = \frac1{\sqrt2}\delta^{\check A}_\alpha,
 \quad
 \bar{\check\xi}^{\dot\alpha\check A} =
 \frac12v_m(\bar\sigma^m)^{\dot\alpha\alpha}\check\xi^{\check A}_\alpha.
\end{equation}
Therefore the $SU(2)_\text{R}$ indices are identified with dotted spinor
indices as before, whereas the $SU(2)_{\check{\text R}}$ indices are
identified as undotted spinor indices.

More generally, starting from a topologically twisted theory on a manifold
with an isometry generated by a Killing vector field $v$, one can
introduce Omega-deformation by choosing the Killing spinor as
(\ref{HO:eq:kso}) and the background fields as in (\ref{HO:eq:bgo}).

\paragraph{The sphere and ellipsoids}

Here we review the construction of a supersymmetric ellipsoid background
following \cite{HO:Hama:2012bg}. The ellipsoid of our interest is defined
as a hypersurface embedded in the flat $\mathbb R^5$,
\begin{equation}
  \frac{x_0^2}{r^2}
 +\frac{x_1^2+x_2^2}{\ell^2}
 +\frac{x_3^2+x_4^2}{\tilde\ell^2}=1.
\label{HO:eq:ell}
\end{equation}
with $U(1)\times U(1)$ isometry. Note that here we are interested
in the ``physical'' SUSY and not the SUSY of topologically twisted
theories, so that the observables should depend non-trivially on some of
the axis-length parameters $\ell,\tilde\ell,r$.
The square of the SUSY will include a linear combinations of the two
$U(1)$ isometries rotating the $12$- and $34$-planes about the origin.

A convenient set of coordinates is the polar angles
$(\rho,\theta,\varphi,\chi)$ which are related to the Cartesian
coordenates on $\mathbb R^5$ as
\begin{align}
 x_0 &\,=\, r\cos\rho, \nonumber \\
 x_1 &\,=\, \ell\sin\rho\cos\theta\cos\varphi, \nonumber \\
 x_2 &\,=\, \ell\sin\rho\cos\theta\sin\varphi, \nonumber \\
 x_3 &\,=\, \tilde\ell\sin\rho\sin\theta\cos\chi, \nonumber \\
 x_4 &\,=\, \tilde\ell\sin\rho\sin\theta\sin\chi.
\label{HO:eq:plc}
\end{align}
The two $U(1)$ isometries of the ellipsoid are generated by Killing
vectors $\partial_\varphi$ and $\partial_\chi$. The north pole
($\rho=0$) and the south pole ($\rho=\pi$) are the two fixed points of
the isometry. Using the above polar angle cordinate system, we see the
ellipsoid as a squashed $S^3$ (with coordinates $\theta,\varphi,\chi$)
fibred over a segment $0\le\rho\le\pi$.

For the round $S^4$ with $\ell=\tilde\ell=r$ the metric becomes
\begin{equation}
 {\rm d}s^2~=~\ell^2({\rm d}\rho^2+\sin^2\rho\cdot{\rm d}s^2_{(S^3)})
 ~=~ E^1E^1+\cdots+E^4E^4.
\end{equation}
A standard choice for the vielbein one-forms $E^a$ is
\begin{equation}
 E^1 = \ell\sin\rho\cos\theta{\rm d}\varphi,~~
 E^1 = \ell\sin\rho\sin\theta{\rm d}\chi,~~
 E^3 = \ell\sin\rho{\rm d}\theta,~~
 E^4 = \ell{\rm d}\rho.
\end{equation}
Note that $E^1,E^2,E^3$ are proportional to the vielbein on the round $S^3$.
A nice fact about this choice of frames is that one can relate part of
the Killing spinor equation on $S^4$ to that on $S^3$, so that the
independent Killing spinors on $S^4$ are all given by those on $S^3$
multiplied by some functions of $\rho$.
Let us choose the following particular solution,
\begin{equation}
\def\arraystretch{1.2}
\begin{array}{l}
 \xi_{A=1}=\sin\tfrac\rho2\cdot\kappa_+, \\
 \xi_{A=2}=\sin\tfrac\rho2\cdot\kappa_-,
\end{array}
\quad
\begin{array}{l}
 \bar\xi_{A=1}=+i\cos\tfrac\rho2\cdot\kappa_+, \\
 \bar\xi_{A=2}=-i\cos\tfrac\rho2\cdot\kappa_-,
\end{array}
\quad
\kappa_\pm=\frac12\left(\begin{array}{r}
    e^{\frac i2(\pm\varphi\pm\chi-\theta)}\\
\mp e^{\frac i2(\pm\varphi\pm\chi+\theta)}
                 \end{array}\right)\,.
\label{HO:eq:kse}
\end{equation}
The square of the corresponding SUSY includes a rotation
$v=\epsilon(\partial_\varphi+\partial_\chi)$ with
$\epsilon=\ell^{-1}$. The theory near the north pole is thus
approximately the topologically twisted theory on
$\mathbb R^4_{\epsilon,\epsilon}$, whereas the theory near the south
pole is the anti-twisted theory on $\mathbb R^4_{\epsilon,-\epsilon}$,
where the minus sign accounts for the relative orientation flip between
the two polar regions.
It then follows from SUSY localization that, as long as we are
interested in supersymmetric observables, the instantons and
anti-instantons have to be localized at the north and south poles
respectively. Their contributions are thus expressed by products of two
Nekrasov partition functions with $\epsilon_1=\epsilon_2=\ell^{-1}$
\cite{HO:Pestun:2007rz}.

It is natural to ask whether there are supersymmetric deformations of
the round sphere geometry which approach the general Omega background,
with $\epsilon_1$ and $\epsilon_2$ independent, near the two poles. A
reasonable guess would be that there should a supersymmetric ellipsoid
background with nonzero auxiliary fields in gravity multiplet, such that
(\ref{HO:eq:kse}) remains a Killing spinor. If that is the case, the
Killing vector field $v$ appearing in the square of supersymmetry is
\begin{equation}
 v ~\equiv~ 2\bar\xi^A\bar\sigma^m\xi_A\partial_m
 ~=~ \epsilon_1\partial_\varphi+\epsilon_2\partial_\chi,
 \quad
 \Big(\epsilon_1\equiv\frac1\ell,~\epsilon_2\equiv\frac1{\tilde\ell}\Big)
\end{equation}
which indeed approach the desired rotation generator near the poles.

It was shown in \cite{HO:Hama:2012bg} that the above naive guess is
actually right. The generalized Killing spinor equation
(\ref{HO:eq:ksh}), with the above form of $\xi_A$ and $\bar\xi_A$
assumed, can be regarded as a linear algebraic equation for the
auxiliary fields $T_{mn},\bar T_{mn},(V_{m})^A_{~B}, M$ in gravity
multiplet. Though the set of equations looks highly over-determined, it
was shown to have a family of solutions. The explicit form of the
background fields was obtained in \cite{HO:Hama:2012bg}. The square of the
supersymmetry was shown to be given by
\begin{equation}
 \HOsusy^2=i{\cal L}_v+\text{Gauge}\big[\hat\Phi\big]
 +\text{R}_{SU(2)}\big[\Theta^A_{~B}\big]
 +\check{\text{R}}_{SU(2)}\big[\check\Theta^{\check A}_{~\check B}\big],
 \quad
 \Theta=\check\Theta=-\frac{\epsilon_1+\epsilon_2}2{\boldsymbol\tau^3}
\label{HO:eq:sye}
\end{equation}
where we used the standard solution for
$\check\xi_{\check A},\bar{\check\xi}_{\check A}$ (\ref{HO:eq:cxi}) to
fix the gauge for local $SU(2)_{\check{\text R}}$ symmetry, and
$\hat\Phi$ was defined in (\ref{HO:eq:htp}).

It is an interesting exercise to study the behavior of the supersymmetric
ellipsoid background near the poles. Near the north pole one can use the
Cartesian coordinates
\begin{equation}
 x_1=\ell\rho\cos\theta\cos\varphi,~~
 x_2=\ell\rho\cos\theta\sin\varphi,~~
 x_3=\tilde\ell\rho\sin\theta\cos\chi,~~
 x_4=\tilde\ell\rho\sin\theta\sin\chi,
\end{equation}
assuming $|x|\ll\ell\sim\tilde\ell$. There the chiral component $\xi_A$
of the Killing spinor (\ref{HO:eq:ksh}) vanishes linearly in $\rho$,
whereas the anti-chiral component $\bar\xi_A$ stays finite. Therefore,
by a suitable local Lorentz and $SU(2)_\text{R}$ rotations it can be
transformed into the form (\ref{HO:eq:kso}). Using
$\epsilon_1=\ell^{-1}$ and $\epsilon_2=\tilde\ell^{-1}$, one can show
the auxiliary field $T_{mn}$ agrees with (\ref{HO:eq:bgo}) and
$\bar T_{mn}=0$ to the leading order in small $\epsilon_i|x|$. However,
the ellipsoids have nonvanishing curvature tensor, and accordingly the
$SU(2)_\text{R}$ gauge field is also non-vanishing. The nonzero
components of the Riemann tensor $R^{ab}_{mn}$ and the $SU(2)_\text{R}$
gauge field strength $(V_{mn})^A_{~B}$, measured in Cartesian
coordinates, are of the order $\epsilon_i^2$. See \cite{HO:Pestun:2014mja}
for the full details.

\paragraph{Local $T^2$-bundle fibrations}

It was shown in \cite{HO:Pestun:2014mja} that the ellipsoid backgrounds of
\cite{HO:Hama:2012bg} can be regarded as an example of supersymmetric {\it
local $T^2$-bundle fibrations}, for which one can apply the same
procedure as explained above to determine the necessary background
auxiliary fields for general squashing parameters.

\section{Partition function}\label{HO:sec:pfn}

Let us review here the application of localization principle to
${\cal N}=2$ supersymmetric path integrals on $S^4$, with some close
look into the use of index theorem and the fixed point formula. We also
present a closed form for the partition function, and review how it
simplifies to a Gaussian matrix integral for ${\cal N}=2^*$ theories for
special choices of mass parameter.

\subsection{Localization principle}\label{HO:sec:loc}

Let us recall how the SUSY localization principle simplifies the
problems of path integration. Suppose a quantum field theory with an
action $S$ and a path-integral measure $\int$ has a supersymmetry $\HOsusy$,
which means that the expectation values of $\HOsusy$-exact observables all
vanish.
\begin{equation}
 \left\langle\HOsusy{\cal O}\right\rangle =
 \int e^{-S}\cdot\HOsusy{\cal O} =0.
\end{equation}
In such a theory, expectation values of $\HOsusy$-invariant observable
are invariant under any deformation of the action of the form $S\to S+t\HOsusy V$,
where the parameter $t$ is arbitrary and $\HOsusy^2V=0$. It is standard to
construct $V$ as the bilinear of all the fermions $\Psi$ and their
$\HOsusy$-variations, because $\HOsusy V$ will then have manifestly
positive-definite bosonic part.
\begin{align}
 V &\,=\, \int {\rm d}^4x\sqrt{g}\sum_\Psi(\HOsusy\Psi)^\dagger\Psi,\nonumber \\
 \HOsusy V &\,=\,
 \int {\rm d}^4x\sqrt{g}\bigg[
   \sum_\Psi(\HOsusy\Psi)^\dagger\HOsusy\Psi+\cdots
   \bigg]\,.
\end{align}
The values of supersymmetric observables should be $t$-independent,
so one may evaluate them at a very large $t$. There the deformed action
is dominated by the term $\HOsusy V$, and nonzero contribution to the path
integral arise only from the vicinity of saddle points characterized by
\begin{equation}
 \HOsusy\Psi=0~~\text{for all the fermions }\Psi.
\label{HO:eq:sdl}
\end{equation}

Let us apply this to the general ${\cal N}=2$ gauge theories of vector
and hypermultiplets on ellipsoids. It is easy to check that the saddle
point equation (\ref{HO:eq:sdl}) is solved by
\begin{align}
\text{vector multiplet}&~:~
 A_m=0,\quad
 \phi=\bar\phi=-\frac i2a_0 ~~\text{(constant)},\quad
 D_{AB}=-ia_0w_{AB},
 \nonumber \\
\text{hypermultiplet}&~:~
 q_A=F_{\check A}=0\,,
\label{HO:eq:spv}
\end{align}
where $w_{AB}$ was introduced in (\ref{HO:eq:wab}). What is more non-trivial is
to prove there are no other saddle points: this has been done explicitly
only for the case of round $S^4$ \cite{HO:Pestun:2007rz}. Assuming that it
continues to be the case for more general ellipsoid backgrounds, one can argue
that the path integral reduces to a finite-dimensional integral over the
space of saddle points parametrized by a Lie algebra-valued constant $a_0$.

An important subtlety in solving the saddle point equation is that, if
one relaxes the condition that the solution be smooth everywhere, the
gauge field is allowed to take nonzero singular values localized
at the two poles \cite{HO:Pestun:2007rz}. The field strength must be
anti-self-dual at the north pole and self-dual at the south pole. This
is how the (anti-)instanton can make nonperturbative contribution to
supersymmetric observables. As was explained in the previous section,
their contribution is precisely given by Nekrasov's partition function,
with argument $q$ for instantons at the north pole and $\bar q$
for the anti-instantons at the south pole.

Localization principle thus leads to the following formula for partition
function,
\begin{equation}
 Z  = \int {\rm d}^ra_0\; e^{-S_\text{cl}(a_0)}
 Z_\text{1-loop}(a_0,m,\epsilon_1,\epsilon_2)
 Z_\text{inst}(a_0,m,q,\epsilon_1,\epsilon_2)
 Z_\text{inst}(a_0,m,\bar q,\epsilon_1,\epsilon_2)\,.
\label{HO:eq:zlc}
\end{equation}
Here the identification $\epsilon_1=1/\ell$, $\epsilon_2=1/\tilde\ell$ was
used. $S_\text{cl}(a_0)$ is the original action evaluated at saddle points, and
the product of Nekrasov's partition function $Z_\text{inst}$ expresses
the contribution of (anti-)instantons at the poles. The one-loop factor
$Z_\text{1-loop}$ arises from path integrating over all the modes
orthogonal to the saddle point locus, for which Gaussian approximation
gives an exact answer thanks to localization principle. Finally,
although the saddle points are labeled by a Lie-algebra valued parameter
$a_0$, the integral can be reduced to its Cartan subalgebra. As is well
known, the invariant measure $[{\rm d}a_0]$ on a Lie algebra is
related to the measure ${\rm d}^ra_0$ on its Cartan subalgebra by
\begin{equation}
 [{\rm d}a_0]~=~ {\rm d}^ra_0\cdot\prod_{\alpha\in\Delta_+}
 (a_0\cdot\alpha)^2\,.
\label{HO:eq:mes}
\end{equation}
In the formula (\ref{HO:eq:zlc}), the Vamdermode factor is understood to
be contained in $Z_\text{1-loop}$.

The SUSY invariant action in general consists of the Yang-Mills term
(\ref{HO:eq:lym}), the Feyet-Illiopoulos term (\ref{HO:eq:lfi}) and the
hypermultiplet kinetic term (\ref{HO:eq:lmt}). Its classical value at the
saddle point $a_0$ is therefore given by the sum of the following,
\begin{equation}
 S_\text{YM}= \frac{8\pi^2}{g^2}\ell\tilde\ell\text{Tr}(a_0^2),\quad
 S_\text{FI}=-16i\pi^2\ell\tilde\ell\zeta a_0,\quad
 S_\text{mat}=0.
\end{equation}
In fact, one can show that $S_\text{mat}$ is exact under the
supersymmetry corresponding to Killing spinors satisfying
$\xi^A\xi_A-\bar\xi^A\bar\xi_A=1$.

\subsection{Gauge fixing}\label{HO:sec:gfx}

We now turn to the explicit path-integration. The first thing we
have to do is to fix a gauge. Following the standard prescription, we
introduce the ghost $c$, anti-ghost $\bar c$ and a Lagrange multiplier
boson $B$. We also introduce a nilpotent symmetry $\HObrs$ which acts on
every physical field $X$ as a gauge transformation by parameter $c$,
\begin{equation}
 \HObrs X = \text{Gauge}[c]X.\quad
\Big(
 \text{example:}~~
 \HObrs A_m=D_mc,~
 \HObrs\lambda_A=i\{c,\lambda_A\}
\Big)
\end{equation}
To achieve nilpotency, the ghost fields should transform by $\HObrs$ as follows,
\begin{equation}
 \HObrs c = icc,\quad
 \HObrs\bar c=B,\quad
 \HObrs B=0.
\end{equation}
Here we decide {\it not} to fix the coordinate-independent part of the
gauge symmetry by this procedure. Therefore the fields $c,\bar c,B$ are
assumed to have no constant modes. (One could alternatively eliminate
the constant modes of those fields by introducing constant
``ghost-for-ghost'' fields \cite{HO:Pestun:2007rz,HO:Hama:2012bg}.) We also
define the action of $\HOsusy$ on the ghost fields
\begin{equation}
 \HOsusy c=a_0-\hat\Phi,\quad
 \HOsusy\bar c=0,\quad
 \HOsusy B=i{\cal L}_v\bar c+i[a_0,\bar c],
\end{equation}
so that the square of the total supercharge $\hat\HOsusy\equiv\HOsusy+\HObrs$
acts on all the fields as follows (compare with the formula
(\ref{HO:eq:sye}) for $\HOsusy^2$),
\begin{equation}
 \hat\HOsusy^2=i{\cal L}_v+\text{Gauge}\big[a_0\big]
 +\text{R}_{SU(2)}\big[\!-\!\tfrac12(\epsilon_1+\epsilon_2)
  \boldsymbol\tau^3\big]
 +\check{\text{R}}_{SU(2)}
  \big[\!-\!\tfrac12(\epsilon_1+\epsilon_2)\boldsymbol\tau^3\big]\,.
\end{equation}
Usual gauge fixing proceeds by choosing an arbitrary gauge-fixing
functional $G$, for example the Lorentz gauge $G=\partial_mA^m$, and
modifying the action by the addition of gauge-fixing term $\HObrs (\bar cG)$.
As was shown in \cite{HO:Pestun:2007rz}, one can replace the gauge-fixing
term by $\hat\HOsusy(\bar cG)$ without changing the value of partition
function. The total supersymmerty $\hat\HOsusy$ is then preserved and can
be used for localization argument.

\subsection{Determinants and index}\label{HO:sec:det}

We now turn to the gauge-fixed path integral with respect to
fluctuations around saddle points. We take the $\hat\HOsusy$-exact deformation
term (including the gauge-fixing term)
\begin{equation}
 \hat\HOsusy\hat V~=~\hat\HOsusy(V+\bar cG),
\label{HO:eq:tdf}
\end{equation}
to be very large, so that Gaussian approximation becomes actually exact
and path integral simply gives rise to determinants.
We also notice that, after the introduction of ghost fields, the number
of bosons and fermions agree off-shell: a vector multiplet consists of
ten bosons $(A_m,\phi,\bar\phi,D_{AB},B)$ and ten fermions
$(\lambda_{\alpha A},\bar\lambda^{\dot\alpha}_A,c,\bar c)$, likewise
a hypermultiplet consists of four bosons $(q_A, F_{\check A})$ and four
fermions $(\psi_\alpha,\bar\psi^{\dot\alpha})$. This is of course
important for the localization principle to work.

We move to a new set of fields which is particularly useful for
evaluating the fluctuation determinant. We first define fermions without spinor
indices from gauginos,
\begin{equation}
 \Psi \equiv
  -i\xi^A\lambda_A-i\bar\xi^A\bar\lambda_A,~~
 \Psi_m \equiv
 i\xi^A\sigma_m\bar\lambda_A-i\bar\xi^A\bar\sigma_m\lambda_A,~~
 \Xi_{AB} \equiv 2\bar\xi_{(A}\bar\lambda_{B)}-2\xi_{(A}\lambda_{B)},
\end{equation}
so that the supersymmetry transformation rule simplifies.
\begin{equation}
 \HOsusy\phi_2 = \Psi,~~
 \HOsusy A_m = \Psi_m,~~
 \HOsusy \Xi_{AB} = D_{AB}+(\cdots).
\end{equation}
Likewise, from the fermion in hypermultiplet we define
\begin{equation}
\def\arraystretch{1.2}
\begin{array}{rcl}
 \Psi_A &\equiv& -i\xi_A\psi+i\bar\xi_A\bar\psi,\\
 \Xi_{\check A} &\equiv&
   \check\xi_{\check A}\psi
 -\bar{\check\xi}_{\check A}\bar\psi,
\end{array}
\quad
\begin{array}{rcl}
 \HOsusy q_A &=& \Psi_A, \\
 \HOsusy\Xi_{\check A} &=& F_{\check A}+(\cdots).
\end{array}
\end{equation}
It is then convenient to take five bosons
${\bf X}=(A_m,\phi_2\equiv\phi-\bar\phi)$, five fermions
$\boldsymbol\Xi=(\Xi_{AB},\bar c,c)$ and their $\hat\HOsusy$-superpartners
as independent variables for vector multiplet. Similarly, for
hypermultiplet we take two bosons ${\bf X}=q_A$, two fermions
$\boldsymbol\Xi=\Xi_{\check A}$ and their $\hat\HOsusy$-superpartners as
independent variables.

In quadratic approximation, the $\hat\HOsusy$-exact deformation term
(\ref{HO:eq:tdf}) decomposes into vectormultiplet and hypermultiplet
parts, and each term has the structure
\begin{align}
 \hat V\Big|_\text{quad.} &\;=\; (\hat\HOsusy{\bf X},{\boldsymbol\Xi})
 \left(\begin{array}{cc}
  D_{00} & D_{01} \\
  D_{10} & D_{11}
       \end{array}\right)
 \left(\begin{array}{c}
  {\bf X} \\ \hat\HOsusy{\boldsymbol\Xi}
       \end{array}\right)
 \nonumber \\
 \hat\HOsusy\hat V\Big|_\text{quad.} &\;=\;
 ({\bf X},\hat\HOsusy{\boldsymbol\Xi})
 \underbrace{
 \left(\begin{array}{cc}
  -{\bf H} & 0 \\
  0 & 1
       \end{array}\right)
 \left(\begin{array}{cc}
  D_{00} & D_{01} \\
  D_{10} & D_{11}
       \end{array}\right)
 }_{= K_\text{b}}
 \left(\begin{array}{c}
  {\bf X} \\ \hat\HOsusy{\boldsymbol\Xi}
       \end{array}\right)
\nonumber \\ &\hskip5mm
-(\hat\HOsusy{\bf X},{\boldsymbol\Xi})
 \underbrace{
 \left(\begin{array}{cc}
  D_{00} & D_{01} \\
  D_{10} & D_{11}
       \end{array}\right)
 \left(\begin{array}{cc}
  1 & 0 \\
  0 & {\bf H}
       \end{array}\right)
 }_{= K_\text{f}}
 \left(\begin{array}{c}
  \hat\HOsusy{\bf X} \\ {\boldsymbol\Xi}
       \end{array}\right),
\end{align}
where we denoted ${\bf H}=\hat\HOsusy^2$.
The Gaussian integral thus gives the square root of the following ratio
of determinants,
\begin{equation}
 \frac{\text{det}K_\text{f}}{\text{det}K_\text{b}}
 = \frac{\text{det}_{\boldsymbol\Xi}{\bf H}}
        {\text{det}_{\bf X}{\bf H}}
 = \frac{\text{det}_{\text{Coker}D_{10}}{\bf H}}
        {\text{det}_{\text{Ker}D_{10}}{\bf H}}\,.
\end{equation}
The last equality follows from the fact that the fields ${\bf X}$
and ${\boldsymbol\Xi}$ take values on the spaces related by the operator
$D_{10}$, and that ${\bf H}$ commutes with $D_{10}$. The ratio of
determinants is closely related to the index defined by
\begin{equation}
 \text{Ind}(D_{10})\equiv
 \text{Tr}_{\text{Ker}D_{10}}\big(e^{-i{\bf H}t}\big)
-\text{Tr}_{\text{Coker}D_{10}}\big(e^{-i{\bf H}t}\big).
\end{equation}

The index can be evaluated using the fixed-point formula, which is
based on the following simple idea. We are interested in the trace of
the operator $e^{-i{\bf H}t}$ involving a finite diffeomorphism
$x^m\to \tilde x^m$, and the index is the difference of the traces
evaluated at the space of fields ${\bf X}$ and ${\boldsymbol\Xi}$. Since
the trace of a matrix is the sum of diagonal elements, the trace of a
finite diffeomorphism operator should be expressed as a ${\rm d}^4x$
integral of a function involving $\delta^4(\tilde x-x)$.
The index is thus expressed as a sum over fixed point contributions,
\begin{equation}
 \text{Ind}(D_{10})
 ~=~ \sum_{x_0:\text{fixed points}}
  \frac{\text{Tr}_{\bf X}(e^{-i{\bf H}t})|_{x_0}-
        \text{Tr}_{\boldsymbol\Xi}(e^{-i{\bf H}t})|_{x_0}}
       {\text{det}(1-\partial\tilde x/\partial x)}\,.
\end{equation}
Defining $z_1=x_1+ix_2$ and $z_2=x_3+ix_4$ from the local Cartesian
coordinate near the poles, one can express the action of
$e^{-i{\bf H}t}$ as
\begin{equation}
 \tilde z_1=z_1q_1=z_1e^{\frac{it}\ell},\quad
 \tilde z_2=z_2q_2=z_1e^{\frac{it}{\tilde\ell}}.
\end{equation}
The determinant in the denominator is therefore given by
\begin{equation}
 \text{det}(1-\partial\tilde x/\partial x)=|(1-q_1)(1-q_2)|^2.
\end{equation}
The enumerator is the difference of the trace of $e^{-i{\bf H}t}$ acting
on fields ${\bf X}$ and ${\boldsymbol\Xi}$ at fixed points. For vector
multiplet fields at the north pole it becomes
\begin{eqnarray}
\lefteqn{
 \text{Tr}_{\bf X}(e^{-i{\bf H}t})|_\text{NP}-
        \text{Tr}_{\boldsymbol\Xi}(e^{-i{\bf H}t})|_\text{NP}
 } \nonumber \\[-4mm]
 \nonumber \\ &=&
 \text{Tr}_\text{adj}(e^{a_0t})\times\Big\{
 (\underbrace{q_1+q_2+\bar q_1+\bar q_2}_{A_m}+
      \underset{\raisebox{-1.3ex}[1ex][0ex]{${}_{\phi_2}$}}{1})
    -(\underbrace{q_1q_2+1+\bar q_1\bar q_2}_{D_{AB}}
     +\underbrace{1+1}_{\bar c,c})
 \Big\}
\end{eqnarray}
The contribution to the index from the North pole is therefore
\begin{equation}
\text{Ind} (D_{10})|_\text{NP}
 ~=~ \text{Tr}_\text{adj}(e^{a_0 t})\times
 \frac{(q_1+q_2+\bar q_1+\bar q_2+1)-(q_1q_2+1+\bar q_1\bar q_2+1+1)}
      {(1-q_1)(1-q_2)(1-\bar q_1)(1-\bar q_2)}\,.
\label{HO:eq:ixn}
\end{equation}
Neglecting some fields whose contribution is trivial, one can identify
the above result with the index of the self-dual complex
($D_\text{SD}:\Omega^0\stackrel{d}\to\Omega^1\stackrel{d^+}\to\Omega^{2+}$)
valued in the adjoint representation of the gauge group, defined
by the instanton equation and gauge equivalence.

If the four factors in the denominator were all expanded into
geometric series, the result would be interpreted as the trace of
$e^{-i{\bf H}t}$ evaluated by expanding all the fields into the basis of
monomial functions $z_1^kz_2^l\bar z_1^m\bar z_2^n$. However, such a
trace would not make sense because there would be infinitely many degenerate
eigenmodes for each eigenvalue of ${\bf H}$. The index does not suffer
from the problem of infinite degeneracy, because the fraction on the
right hand side of (\ref{HO:eq:ixn}) is reducible reflecting the
cancellation between the fields ${\bf X}$ and ${\boldsymbol\Xi}$. But
there remains another more subtle issue which requires a careful
regularization, as we will see below.

After simplifying the fraction, combining the contributions from the two
poles and recalling that the fields $c,\bar c$ do not have constant
modes, the index is given by
\begin{equation}
 \text{Ind}(D_{10})|_\text{vec}=
 \text{Tr}_\text{adj}(e^{a_0 t})\times\left\{
  \Big[-\frac{1+q_1q_2}{(1-q_1)(1-q_2)}\Big]_\text{NP}
 +\Big[-\frac{1+q_1q_2}{(1-q_1)(1-q_2)}\Big]_\text{SP}+2
 \right\}
\end{equation}
for vector multiplet. Similarly, for hypermultiplet in the
represetentation $R$ of the gauge group the index becomes
\begin{equation}
 \text{Ind}(D_{10})_\text{hyp}=
 \text{Tr}_{{R}+\bar{R}}(e^{a_0 t})\times\left\{
  \Big[\frac{(q_1q_2)^{\frac12}}{(1-q_1)(1-q_2)}\Big]_\text{NP}
 +\Big[\frac{(q_1q_2)^{\frac12}}{(1-q_1)(1-q_2)}\Big]_\text{SP}
 \right\}.
\end{equation}
To read from the index the spectrum of ${\bf H}$ which is necessary for
the computation of one-loop determinant, one needs to expand the above
expressions into series in $q_1,q_2$. But a priori there is no natural
choice whether to expand in positive or negative series in $q$'s. We have seen
above that fixed point formula allows one to express the index as a sum
of pole contributions, but it does not give us any further information about
which eigenmode of ${\bf H}$ is supported around which pole. Indeed,
although the index of a differential operator $D_{10}$ depends only on
the term of highest order in the derivative, the detailed behavior of
its zeromodes depends on the subleading terms as well. One can choose
the subleading term in any convenient manner so that each eigenmode of
${\bf H}$ has localized support near one of the poles. The index
should of course be independent of such regularizations.

Let us look into this point in more detail, taking the hypermultiplet index
as an example. To the leading order in the derivatives, the differential
operator $D_{10}$ is given by
\begin{equation}
 \Xi^{\check A} (D_{10})_{\check AB} q^B ~=~
 \Xi^{\check A}\big(i\bar{\check\xi}_{\check A}\bar\sigma^m\xi_B
               -i\check\xi_{\check A}\sigma^m\bar\xi_B\big)D_mq^B.
\end{equation}
We are interested in how the zeromode wavefunctions get localized near the
poles depending on the choice of the non-derivative terms. Since
$D_{10}$ has to commute with ${\bf H}$, we follow the suggestion in
\cite{HO:Pestun:2007rz} and introduce a non-derivative term in $D_{10}$
through the modification $D_m\to D_m-2isv_m$, where $s$
is an arbitrary real parameter. Similar modification of differential
operators was considered in the study of Morse theory \cite{HO:Witten:1982im}
and in particular the derivation of holomorphic Morse inequality in
\cite{HO:Witten:1984aa}.

Near the north pole one may identify $\Xi^{\check A}$ as a chiral spinor
and $q^A$ as an anti-chiral spinor, and $(D_{10})_{\alpha\dot\beta}$ is
then simply the Dirac operator
\begin{equation}
\def\arraystretch{1.2}
 D_{10} = \frac 12\sigma^m(i\partial_m+sv_m)
 = \left(\begin{array}{rr}
   \partial_{\bar z_2}+s\epsilon_2z_2 &
   \partial_{z_1}-s\epsilon_1z_1\\
   \partial_{\bar z_1}+s\epsilon_1z_1 &
  -\partial_{z_2}+s\epsilon_2z_2
	 \end{array}\right)\,.
\label{HO:eq:d10}
\end{equation}
For vector multiplet, the relevant differential operator near the north
pole has the index structure $(D_{10})_{\dot\alpha\dot\beta,\gamma\dot\delta}$,
and is the adjoint of the operator above twisted by an anti-chiral
spinor bundle. Assuming that $\epsilon_1=\ell^{-1}$ and
$\epsilon_2=\tilde\ell^{-1}$ are both positive, the operator $D_{10}$
can be shown to have no $\Xi$-zeromodes, but it has $q$-zeromodes of the
following form,
\begin{align}
 s>0 ~~\Longrightarrow~~
\def\arraystretch{1.2}
 q^A=\left(\begin{array}{c}
	     z_1^mz_2^n e^{-s(\epsilon_1|z_1|^2+\epsilon_2|z_2|^2)}\\0
		  \end{array}\right),&\quad
 e^{-i{\bf H}t}=e^{a_0t}\cdot q_1^{m+\frac12}q_2^{n+\frac12},
 \nonumber \\
 s<0 ~~\Longrightarrow~~
\def\arraystretch{1.2}
 q^A=\left(\begin{array}{c}
	0\\\bar z_1^m\bar z_2^n e^{+s(\epsilon_1|z_1|^2+\epsilon_2|z_2|^2)}
		  \end{array}\right),&\quad
 e^{-i{\bf H}t}=e^{a_0t}\cdot q_1^{-m-\frac12}q_2^{-n-\frac12}.
\end{align}
This indicates one should expand the north-pole contribution to the
index into positive (negative) series in $q_1,q_2$ if $s>0$
(resp. $s<0$). The analysis goes similarly near the south pole, with the
result that one has to series-expand in the opposite way. We thus arrive
at the formula for the index,
\begin{align}
 \text{Ind}(D_{10})|_\text{vec} &\;=\;
  \text{Tr}_\text{adj}(e^{a_0 t})\Big\{2-\sum_{m,n\ge0}
  \big(q_1^mq_2^n+q_1^{m+1}q_2^{n+1}+q_1^{-m}q_2^{-n}+q_1^{-m-1}q_2^{-n-1}
  \big)\Big\},
 \nonumber \\
 \text{Ind}(D_{10})|_\text{hyp} &\;=\;
  \text{Tr}_{R+\bar R}(e^{a_0 t})\sum_{m,n\ge0}\big(
  q_1^{m+\frac12}q_2^{n+\frac12}+q_1^{-m-\frac12}q_2^{-n-\frac12}\big).
\end{align}
Note the operator $D_{10}$ has infinitely many zeromodes, owing to the fact
that it is not elliptic but only transversely elliptic \cite{HO:Atiyah:1974aa}.

The one-loop determinant factor $Z_\text{1-loop}$ in (\ref{HO:eq:zlc}) can
be easily obtained from the above formula for the index. We assume $a_0$
to be in Cartan subalgebra and neglect $a_0$-independent factors.
One then finds that $Z_\text{1-loop}$ is a product of contributions from
vector and hypermultiplets,
\begin{align}
 Z_\text{1-loop}^\text{vec} &\;=\;
 \prod_{\alpha\in\Delta_+}
 \frac{\Upsilon(i\hat a_0\cdot\alpha)\Upsilon(-i\hat a_0\cdot\alpha)}
      {(\hat a_0\cdot\alpha)^2}\times (\hat a_0\cdot\alpha)^2
 \;=\;
 \prod_{\alpha\in\Delta}
 \Upsilon(i\hat a_0\cdot\alpha),
 \nonumber \\
 Z_\text{1-loop}^\text{hyp} &\;=\;
 \prod_{\rho\in R}\Upsilon(\tfrac Q2+i\hat a_0\cdot\rho)^{-1}\,,
\label{HO:eq:zol}
\end{align}
where we included the Vandermonde determinant (\ref{HO:eq:mes}) into
$Z^\text{vec}_\text{1-loop}$. 
Here $\hat a_0=\sqrt{\ell\tilde\ell}a_0$ is the normalized saddle-point
parameter, $\alpha\in\Delta_+$ runs over positive roots of the gauge Lie
algebra and $\rho\in R$ runs over weights of the representation $R$.
The function $\Upsilon(x)$ is defined as an infinite product,
\begin{equation}
 \Upsilon(x)=\text{const}\cdot\prod_{m,n\ge0}(x+mb+nb^{-1})(Q-x+mb+nb^{-1}),
 \quad\Big(Q=b+\frac1b\Big)
\end{equation}
where the parameter $b$ is related to the ellipsoid geometry by
$b=(\ell/\tilde\ell)^{\frac12}$. It appears frequently in observables of
Liouville or Toda CFTs with coupling $b$. See for example
\cite{HO:Zamolodchikov:1995aa}, where some important properties of
$\Upsilon(x)$ are also summarized.

Note that one can read off an information on one-loop running of the
gauge coupling from the behavior of $Z_\text{1-loop}$ for
$\ell\tilde\ell\ll a_0^2$,
\begin{equation}
 S_\text{YM}= \frac{8\pi^2}{g^2}\text{Tr}\big(\hat a_0^2\big),\quad
-\ln Z_\text{1-loop} \sim
  \ln(\ell\tilde\ell)^{\frac12}\cdot
 \Big\{ \text{Tr}_\text{adj}\big(\hat a_0^2\big)
      - \text{Tr}_R\big(\hat a_0^2\big) \Big\}\,.
\end{equation}
Here we used the asymptotic behavior of $\Upsilon(x)$ at large $|x|$,
\begin{equation}
 \ln\Upsilon(x) \sim \Big(x-\frac Q2\Big)^2\ln x
+ \Big(\frac16-\frac{Q^2}{12}\Big)\ln x
 -\frac 32\Big(x-\frac Q2\Big)^2+\cdots.
\end{equation}

\subsection{${\cal N}=4$ SYM and Gaussian matrix model}\label{HO:sec:mm}

${\cal N}=2$ gauge theory with massless adjoint hypermultiplet has an
enhanced supersymmetry and is called ${\cal N}=4$ SYM. Application of
localization principle to this model is particularly interesting since
one can expect to obtain nontrivial and precise evidences for the AdS/CFT
correspondence. In this respect, there was a long standing conjecture
that the expectation value of circular Wilson loops in ${\cal N}=4$ SYM
is given by a simple Gaussian matrix integral
\cite{HO:Erickson:2000af,HO:Drukker:2000rr}. Pestun's work
\cite{HO:Pestun:2007rz} gave an analytic proof of this conjecture.

The ${\cal N}=4$ SYM can be deformed to the so-called ${\cal N}=2^\ast$
theory by making the adjoint hypermultiplet massive. The measure and the
one-loop determinant part of the ellipsoid partition function for
this theory read
\begin{equation}
 Z_\text{1-loop}~=~
 \prod_{\alpha\in\Delta_+}
 \frac{\Upsilon(i\hat a_0\cdot\alpha)
       \Upsilon(-i\hat a_0\cdot\alpha)}
      {\Upsilon(\frac Q2+i\hat m+i\hat a_0\cdot\alpha)
       \Upsilon(\frac Q2+i\hat m-i\hat a_0\cdot\alpha)},
\end{equation}
where $\hat m$ is the normalized (dimensionless) hypermultiplet
mass. Note that it is invariant under sign-flip of $\hat m$ since
$\Upsilon(x)=\Upsilon(Q-x)$.

An obvious special value of the mass is $\hat m=\pm iQ/2$, for which
the $\Upsilon$ functions in the denominator and enumerator cancel
precisely. Similar simplification happens also to Nekrasov's partition
function. For example for $U(N)$ gauge group, $Z_\text{inst}$ is simply
given by a sum over the sets of $N$ Young diagrams weighted by $q^k$,
where $k$ is the total number of boxes in the $N$ diagrams. Therefore
\begin{equation}
 Z_\text{inst}=\prod_{k\ge1}(1-q^k)^{-N}.
\end{equation}
The only $a_0$-dependence remaining in the integrand is the classical
action $S_\text{YM}$. The $a_0$ integral can be easily performed and
gives $(\text{Im}\tau)^{-N/2}$. The result agrees with the torus
partition function of the 2D CFT of $N$ massless scalars, but is different
from Gaussian matrix integral.

Another special value of the mass is $\hat m=\pm\frac i2(b^{-1}-b)$, for
which the measure and the determinant become
\begin{equation}
 Z_\text{1-loop}
 ~=~
 \prod_{\alpha\in\Delta_+}
 \frac{\Upsilon(i\hat a_0\cdot\alpha)
       \Upsilon(-i\hat a_0\cdot\alpha)}
      {\Upsilon(b^{\pm1}+i\hat a_0\cdot\alpha)
       \Upsilon(b^{\pm1}-i\hat a_0\cdot\alpha)}
 ~=~
 \prod_{\alpha\in\Delta_+}(\hat\alpha_0\cdot a)^2\,,
\end{equation}
which is the natural measure for matrix integral. At the same time, the
Nekrasov partition function becomes trivial for this special value of
$\hat m$ , namely $Z_\text{inst}=1$, due to the emergence of fermionic
zeromodes in the moduli space of $k(\ge1)$ instantons. As was argued in
\cite{HO:Okuda:2010ke}, the additional fermion zeromode is the consequence
of supersymmetry enhancement. Thus the SUSY path integral reduces to the
Gaussian matrix integral for this special choice of $\hat m$.

\section{Supersymmetric observables}\label{HO:sec:vev}

We review here the application of localization principle to the
evaluation of supersymmetric non-local observables -- Wilson loops, 't
Hooft loops and surface operators.

\subsection{Wilson loops}\label{HOsec:wsn}

Having understood how to compute partition function using localization
principle, it is straightforward to include Wilson loop
operators. Wilson loops are defined as usual by holonomy
integrals along closed paths, but in supersymmetric Wilson loops the
gauge field is accompanied by scalar fields in vector multiplet. Also,
the loops have to be aligned with the direction of the isometry
generated by $\HOsusy^2$. For generic mutually incommensurable choice of
$\ell,\tilde\ell$, there are only two types of supersymmetric closed paths:
\begin{align}
 S^1_\varphi(\rho) &~:~
 (x_0,x_1,x_2,x_3,x_4)=
 (r\cos\rho,\ell\sin\rho\cos\varphi,\ell\sin\rho\sin\varphi,0,0),
 \nonumber \\
 S^1_\chi(\rho) &~:~
 (x_0,x_1,x_2,x_3,x_4)=
 (r\cos\rho,0,0,\tilde\ell\sin\rho\cos\chi,\tilde\ell\sin\rho\sin\chi).
\end{align}
Namely, $S^1_\varphi(\rho)$ is a circle within an $(x_1,x_2)$-plane at a
fixed $x_0$ and $x_3=x_4=0$, and similarly $S^1_\chi(\rho)$ is a circle
within an $(x_3,x_4)$-plane. The corresponding Wilson loop operators are
\begin{align}
 W_\varphi(R) &\;\equiv\; \text{Tr}_R\text{P}
 \exp i\int_{S^1_\varphi(\rho)}{\rm d}\varphi
 \Big(A_\varphi-2\ell(\phi\cos^2\tfrac\rho2+\bar\phi\sin^2\tfrac\rho2)\Big),
 \nonumber \\
 W_\chi(R) &\;\equiv\; \text{Tr}_R\text{P}
 \exp i\int_{S^1_\chi(\rho)}{\rm d}\varphi
 \Big(A_\chi-2\tilde\ell(\phi\cos^2\tfrac\rho2+\bar\phi\sin^2\tfrac\rho2)\Big).
\end{align}
Note that the integrand is proportional to $\hat\Phi$ of (\ref{HO:eq:htp})
evaluated along the path, so the SUSY invariance is very easy to check.
The expectation values of these operators can thus be evaluated by just
inserting their classical values
\begin{equation}
 W_\varphi(R)= \text{Tr}_R\exp\left(-2\pi b\hat a_0\right),\quad
 W_\chi(R)= \text{Tr}_R\exp\left(-2\pi b^{-1}\hat a_0\right).
\end{equation}
into the integrand of (\ref{HO:eq:zlc}).

\subsection{'t Hooft loops}\label{HO:sec:tht}

't Hooft loops play an equally important role as Wilson loops. They were
originally introduced in \cite{HO:'tHooft:1977hy} as a probe to
distinguish different phases of gauge theories. Also, in 4D ${\cal N}=2$
SUSY gauge theories, the Wilson and 't Hooft loop operators are known to
transform among one another under duality.

\paragraph{Definition of 't Hooft operator}

A 't Hooft operator introduces a Dirac monopole singularity along a
path in a 4D space, and its charge is specified by a coweight $B$ of
the gauge group. Insertions of 't Hooft operators therefore not only
changes the classical SYM action $S_\text{cl}$, but also affects the
one-loop and instanton parts of the formula (\ref{HO:eq:zlc}) since it
changes the boundary condition for the path integration variables. This
problem was analized in detail in \cite{HO:Gomis:2011pf} for a single 't
Hooft operator inserted along a great circle in the equator $S^3$ of the
round sphere.

Let us first study the operator lying along the
$x^1$-axis ($x^2=x^3=x^4=0$) in the flat $\mathbb R^4$. The behavior of
the magnetic field around it is
\begin{equation}
 F \sim -\frac B4\epsilon_{ijk}\frac{x^i{\rm d}x^j{\rm d}x^k}{|x|^3}
 ~~~~(i,j,k=2,3,4).
\end{equation}
When the $\theta$-angle is nonzero, the presence of magnetic charge
changes the quantization condition of electric charge
\cite{HO:Witten:1979ey}. This implies that the 't Hooft operator also
induces nonzero electric field proportional to $\theta$,
\begin{equation}
 F_{1i} \sim \frac{i\theta g^2 B}{16\pi^2}\frac{x^i}{|x|^3}\,.
\end{equation}
If we require the 't Hooft operator to be half-BPS, the scalars are also
required to take non-zero values around it. If the unbroken
supersymmetry is characterized by $\xi_A=\sigma_1\bar\xi_A e^{i\alpha}$,
the scalars have to behave near the 't Hooft operator as follows,
\begin{equation}
 \phi \sim e^{i\alpha}
 \left(\frac14-\frac{i\theta g^2}{32\pi^2}\right)\frac{B}{|x|},
 \quad
 \bar\phi \sim e^{-i\alpha}
 \left(-\frac14-\frac{i\theta g^2}{32\pi^2}\right)\frac{B}{|x|}.
\end{equation}

Cosider now general ${\cal N}=2$ SUSY theories on the round $S^4$ with
radius $\ell$, and put a 't Hooft operator with charge $B$ along the circle
$S^1_\varphi$ at $\rho=\pi/2$, namely the intersection of the sphere
(\ref{HO:eq:ell}) with $x_0=x_3=x_4=0$. Our Killing spinor
(\ref{HO:eq:kse}) satisfies $\xi_A=-\sigma_1\bar\xi_A$ there, so we
substitute $e^{i\alpha}=-1$ into the above expressions for fields on
$\mathbb R^4$ and then map to $S^4$. Using the Cartesian coordinates
$x_0,\cdots,x_4$ introduced in (\ref{HO:eq:ell}), the value of gauge and
scalar fields is
\begin{align}
 F &= -\frac B{4|x|^3}\epsilon_{ijk}x_i{\rm d}x_j {\rm d}x_k
     +\frac{i\theta g^2B}{16\pi^2}\frac{\ell {\rm d}x_1{\rm d}x_2}{|x|^3},
 \qquad(i,j,k=0,3,4)
 \nonumber \\
 \phi &= \left(-\frac14+\frac{i\theta g^2}{32\pi^2}\right)\frac{B}{|x|}
 -\frac{ia_0}2,
 \qquad
 \bar\phi = 
 \left(\frac14+\frac{i\theta g^2}{32\pi^2}\right)\frac{B}{|x|}
 -\frac{ia_0}2.
\label{HO:eq:sdm}
\end{align}
Here we used $|x|\equiv\sqrt{x_0^2+x_3^2+x_4^2}$, and we also included
the constant terms for the scalars. It was shown in \cite{HO:Gomis:2011pf}
that the above expression with $[B,a_0]=0$ exhausts all the saddle point
configurations with the correct singular behavior of fields around the
loop.

\paragraph{Localization computation}

To compute the expectation values of 't Hooft operators, one needs to
work out the classical action on the saddle-point configuration
(\ref{HO:eq:sdm}), one-loop determinant and instanton contribution.
All of them receive nontrivial modification from 't Hooft operator, as
we will now review.

The classical SYM action integral diverges near the 't Hooft loop since it
corresponds to the self-energy of monopole. It can be regularized by
removing a neighborhood of the loop $B_3\times S^1$ from the integration
domain, and adding the boundary term
\begin{equation}
 S_\text{boundary}~=~
 i\ell\int_{S^2\times S^1} \frac{{\rm d}\varphi}{2\pi}\text{Tr}
 \left(e^{-i\alpha}\tau\phi + e^{i\alpha}\bar\tau\bar\phi\right)F.
\end{equation}
Here $\varphi$ is the coordinate along the loop and $\tau$ is the
complexified gauge coupling. The total classical action evaluated on the
saddle point (\ref{HO:eq:sdm}) is thus finite,
\begin{align}
& (S_\text{YM}+S_\text{boundary})\Big|_\text{cl}
 ~=~ -i\pi\tau\text{Tr}(\hat a_\text{N}^2)
  +i\pi\bar\tau\text{Tr}(\hat a_\text{S}^2),
 \nonumber \\
& \hat a_\text{N} \equiv a_0\ell -\frac{\theta g^2B}{16\pi^2}+\frac{iB}2,
 \quad
 \hat a_\text{S} \equiv a_0\ell -\frac{\theta g^2B}{16\pi^2}-\frac{iB}2.
\end{align}
We notice here that $\hat a_\text{N}$ and $\hat a_\text{S}$ are the
values of the scalar $\hat\Phi$ (\ref{HO:eq:htp}) at the two poles, which are
relevant in the evaluation of equivariant integrals over the instanton
moduli spaces there. Therefore the argument of Nekrasov's partition
functions representing the effect of instantons at the north pole
(anti-instantons at the south pole) should be changed from $\hat a_0$ to
$\hat a_\text{N}$ (resp. $\hat a_\text{S}$).

Actually there is a subtlety in identifying
$\hat a_\text{N},\hat a_\text{S}$ with the value of $\hat\Phi$, since
the latter contains the gauge potential $A_m$ and there is no globally
well-defined expression for it in the presence of the 't Hooft operator.
By integrating the expression for the field strength
(\ref{HO:eq:sdm}) one finds
\begin{align}
 & A =
  -\frac B2\Big(\frac{x_0}{|x|}-C\Big){\rm d}\chi
  +\frac{i\theta g^2B}{16\pi^2}\Big(\frac\ell{|x|}-1\Big){\rm d}\varphi,
 \nonumber \\ &
 \bigg(
 {\rm d}\chi=\frac{x_3{\rm d}x_4-x_4{\rm d}x_3}{x_3^2+x_4^2},
 \quad
 {\rm d}\varphi=\frac{x_1{\rm d}x_2-x_2{\rm d}x_1}{x_1^2+x_2^2},
 \quad
 |x|=\sqrt{x_0^2+x_3^2+x_4^2}.
 \bigg)
\end{align}
Near the north and south poles, we choose the integration constant $C$
as $+1$ or $-1$ to avoid Dirac string singularity and find
$\hat\Phi=\hat a_\text{N}$ or $\hat\Phi=\hat a_\text{S}$,
respectively. Near the equator, the natural choice $C=0$ leads to
\begin{equation}
 \hat\Phi=\hat a_\text{E}\equiv a_0\ell -\frac{\theta g^2B}{16\pi^2}\,.
\end{equation}

Let us next turn to the evaluation of one-loop Gaussian integral over
fluctuations from the above saddle points. As in the previous section
one can relate it to an index and express it as a sum over contributions
from fixed points. In addition to the north and south poles, this time
there is a nontrivial contribution from the equator in the vicinity of
the loop, due to the change in the boundary condition of fields
there. We introduce the coordinates $\varphi\sim\varphi+2\pi$ and
$y_1,y_2,y_3$ to parametrize the local geometry $S^1\times \mathbb R^3$
near the loop, assuming the loop is at the origin of $\mathbb R^3$. The
coordinate $\varphi$ here is the same as the angle coordinate $\varphi$
in (\ref{HO:eq:plc}), while the other angle coordinate $\chi$ there
corresponds to the rotation angle in $(y_1,y_2)$-plane here.

Our Killing spinor $\xi_A,\bar\xi_A$ (\ref{HO:eq:kse}) and
$\check\xi_{\check A},\bar{\check\xi}_{\check A}$ are anti-periodic
in $\varphi$. It is convenient to use a local $J_3$ transformations in
$SU(2)_\text{R}$ and $SU(2)_{\check{\text R}}$ to make them all
independent of $\varphi$. Then vector multiplet fields become all periodic
in $\varphi$, while hypermultiplet fields are all antiperiodic.
The index involves the trace of $e^{-it{\bf H}}$, where
\begin{equation}
 {\bf H}=\hat\HOsusy^2 ~=~ \frac 1\ell\Big(
  i\partial_\varphi + i\partial_\chi + \text{Gauge}[\hat a_\text{E}] \Big)
  ~\equiv~ \frac i\ell\partial_\varphi + {\bf H}_{(3)}.
\end{equation}
Kaluza-Klein expansion with respect to $\varphi$ thus relates the
equatorial contribution to the index of our interest to a 3D
index. The reduction takes the following schematic form
\begin{equation}
 \text{Ind}(D_{(4)})\Big|_\text{eq}
 = \sum_{n}e^{\frac{int}\ell}\;\text{Ind}(D_{(3)}).
\end{equation}
The sum with respect to $n$ is over integers for vector multiplet index
and half-odd integers for hypermultiplets. For vector multiplet, the
natural choice for the operator $D_{(3)}^\text{vec}$ is the one
associated with the gauge equivalence classes of small fluctuations
around the singular solution to Bogomolny equation $F + \ast D\phi_2=0$,
\begin{equation}
 F = -\frac B{4|y|^3}\epsilon_{ijk}y_i{\rm d}y_j{\rm d}y_k,\quad
 \phi_2=-\frac12\frac B{|y|}\,.
\label{HO:eq:sgm}
\end{equation}
For hypermultiplet, the natural choice is the 3D Dirac operator
$D_{(3)}^\text{hyp}=i{\boldsymbol\tau}^i(\partial_{y_i}-iA_i)+\phi_2$.

In \cite{HO:Gomis:2011pf} the 3D indices were evaluated by using
Kronheimer's construction of $U(1)$-invariant instantons
\cite{HO:Kronheimer:1986ms}. Consider
Gibbons-Hawking parametrization of flat $\mathbb C^2$,
\begin{align}
 {\rm d}z_1{\rm d}\bar z_1+{\rm d}z_2{\rm d}\bar z_2 &\;=\;
 \frac1{4r}({\rm d}r^2+r^2{\rm d}\vartheta^2+r^2\sin^2\vartheta{\rm d}\chi^2)
 +r({\rm d}\psi-\frac12\cos\vartheta{\rm d}\chi)^2
 \nonumber \\ &\;=\;
 \frac1{4|y|}{\rm d}y_i{\rm d}y_i+|y|({\rm d}\psi+\omega)^2\,.
 \nonumber \\
&\left(
 z_1=\sqrt{r}\cos\tfrac\vartheta2 e^{\frac{i\chi}2-i\psi},\quad
 z_2=\sqrt{r}\sin\tfrac\vartheta2 e^{\frac{i\chi}2+i\psi}
\right)
\end{align}
An important fact here is that, if $(A,\phi_2)$ satisfies Bogomolny
equation on $\mathbb R^3$, then
\begin{equation}
 {\cal A}= A -2|y|\phi_2({\rm d}\psi+\omega),
\end{equation}
is an anti-self-dual and $\psi$-translation invariant gauge field
configuration on $\mathbb C^2$. Note also that the singular monopole
solution (\ref{HO:eq:sgm}) corresponds to a pure gauge ${\cal A}=B{\rm d}\psi$
under this map. This map also relates the 3D indices of interest to the
restricted 4D indices, where the trace is taken only over the space of
$\psi$-independent wave functions. For example,
the index of $D^\text{vec}_{(3)}$ can be computed from the
index of 4D self-dual complex $D_\text{SD}$ associated to the gauge
equivalence class of fluctuations from an anti-self-dual connection ${\cal A}$,
restricted to $\psi$-independent wave functions. The 3D index
is thus obtained by avaraging the 4D index over $\psi$-translations
\begin{eqnarray}
 \text{Ind}(D_{(3)}^\text{vec})
 &=&
 \int_0^{2\pi}\frac{{\rm d}\nu}{2\pi}
 \left\{
 \text{Tr}_{\text{Ker}D_\text{SD}}
 (e^{-it{\bf H}_{(3)} +\nu\partial_\psi})
-\text{Tr}_{\text{Coker}D_\text{SD}}
 (e^{-it{\bf H}_{(3)} +\nu\partial_\psi})
 \right\}_{{\cal A}=B{\rm d}\psi}
 \nonumber \\
 &=&
 \int_0^{2\pi}\frac{{\rm d}\nu}{2\pi}
 \left\{
 \text{Tr}_{\text{Ker}D_\text{SD}}
 (e^{-it{\bf H}_{(3)} +\nu(\partial_\psi+iB)})
-\text{Tr}_{\text{Coker}D_\text{SD}}
 (e^{-it{\bf H}_{(3)} +\nu(\partial_\psi+iB)})
 \right\}_{{\cal A}=0}
 \nonumber \\ &=&
 \int_0^{2\pi}\frac{{\rm d}\nu}{2\pi}\;
 \frac{\text{Tr}_\text{adj}(e^{\hat a_\text{E}t+i\nu B})\times
       (1+e^{-\frac {it}\ell})
       (1-e^{\frac {it}{2\ell}+i\nu})
       (1-e^{\frac {it}{2\ell}-i\nu})}
      {(1-\delta e^{\frac {it}{2\ell}-i\nu})
       (1-\delta e^{\frac {it}{2\ell}+i\nu})
       (1-\delta e^{-\frac {it}{2\ell}-i\nu})
       (1-\delta e^{-\frac {it}{2\ell}+i\nu})}.
\end{eqnarray}
Here in the second line we similarity-transform all the operators
involved by a gauge rotation, and in the last line we introduced a
parameter $\delta$ ($0<\delta<1$ and $\delta\to 1$) to indicate
expansions into geometric series. The index of $D_{(3)}^\text{hyp}$ is
related to the 4D Dirac index in the same way.
The final result is
\begin{eqnarray}
 \text{Ind}(D_{(3)}^\text{vec}) &=&
 -\frac12(u+u^{-1})\sum_{\alpha\in\Delta_+}
 (e^{\alpha\cdot\hat a_\text{E}t}+e^{-\alpha\cdot\hat a_\text{E}t})
 \frac{u^{|\alpha\cdot B|}-u^{-|\alpha\cdot B|}}{u-u^{-1}},
 \nonumber \\
 \text{Ind}(D_{(3)}^\text{hyp}) &=&
 \frac12\sum_{\rho\in R}
 (e^{\rho\cdot\hat a_\text{E}t-\hat mt}
 +e^{-\rho\cdot\hat a_\text{E}t+\hat mt})
 \frac{u^{|\rho\cdot B|}-u^{-|\rho\cdot B|}}{u-u^{-1}}\,.
\end{eqnarray}
Here we used $u\equiv e^{it/2\ell}$, and $\hat m$ is the normalized mass
parameter for the hypermultiplet. Note also that by definition of
coweight $B$ the inner products $\alpha\cdot B$ and $\rho\cdot B$ are
always integers.

Let us now present the formula for the expectation value of a 't Hoof
operator $T_B$. Without loss of generality we can choose the charge $B$
to be the highest weight vector of an irreducible representation of $^LG$
(Langlands dual of the gauge group). For ``small'' charge $B$, all the
weight vectors of the corresponding representation are Weyl images of
$B$. In such cases, the expectation value of the 't Hooft operator can
be expressed by combining all the arguments reviewed above,
\begin{align}
\left\langle T_B\right\rangle
\;=\;
 \int [{\rm d}\hat a_\text{E}]\;&
 q^{\frac12\text{Tr}(\hat a_\text{N})^2}
 Z_\text{1-loop}(\hat a_\text{N},\hat m)^{\frac12}
 Z_\text{inst}(\hat a_\text{N},\hat m,q)
 \cdot Z_\text{1-loop}^\text{(eq)}(\hat a_\text{E},\hat m,B)
\nonumber \\ \cdot\, &
 \bar q^{\frac12\text{Tr}(\hat a_\text{S})^2}
 Z_\text{1-loop}(\hat a_\text{S},\hat m)^{\frac12}
 Z_\text{inst}(\hat a_\text{S},\hat m,\bar q)
\nonumber \\ &
 \hat a_\text{N}=\hat a_\text{E}+\frac{iB}2,\quad
 \hat a_\text{S}=\hat a_\text{E}-\frac{iB}2\,.
\label{HO:eq:tbg}
\end{align}
This can be rewritten further as a sum over Weyl images of $B$. As an
example, consider $SU(N)$ ${\cal N}=2^\ast$ theory on $S^4$ and take as
$B$ the highest weight vector for fundamental representation, namely
$B=h_1=(\frac{N-1}N,-\frac1N,\cdots,-\frac1N)$.
Then the eqpectation value is expressed as a sum over weight vectors
$h_k$,
\begin{align}
\left\langle T_B\right\rangle
\;=\;\frac1N\sum_{k=1}^N
 \int {\rm d}^r\hat a\;&
 q^{\frac12\text{Tr}(\hat a+\frac i2h_k)^2}
 Z_\text{1-loop}(\hat a+\tfrac i2h_k,\hat m)^{\frac12}
 Z_\text{inst}(\hat a+\tfrac i2h_k,\hat m,q)
 \,Z_\text{1-loop}^\text{(eq)}(\hat a,\hat m,h_k)
\nonumber \\ \cdot\,&
 \bar q^{\frac12\text{Tr}(\hat a-\frac i2h_k)^2}
 Z_\text{1-loop}(\hat a-\tfrac i2h_k,\hat m)^{\frac12}
 Z_\text{inst}(\hat a-\tfrac i2h_k,\hat m,\bar q)\,,
\label{HO:eq:tbf}
\end{align}
where the one-loop determinant factor from the equator is
\begin{equation}
 Z_\text{1-loop}^{(\text{eq})}(\hat a,\hat m,h_k)=
 \prod_{j\ne k}
 \frac{\{\cosh\pi(\hat a_k-\hat a_j+\hat m)
        \cosh\pi(\hat a_k-\hat a_j-\hat m)\}^{\frac12}}
      {\cosh\pi(\hat a_k-\hat a_j)}.
\end{equation}
This result was shown to agree with the expectation value of Verlinde's
loop operators in $A_{N-1}$ Toda CFT.

\paragraph{Monopole screening}

As in (\ref{HO:eq:tbf}), the expectation value of a 't Hooft operator
$\langle T_B\rangle$ in $S^4$ for general magnetic charge $B$ involves
the sum over weight vectors $h$ of the highest weight representation
$B$ of the group $^LG$. The weight vector $h$ appearing in the argument of
$Z_\text{1-loop}$ and $Z_\text{inst}$ has an interpretation as
the value of magnetic charge measured at the polar regions. Now for a
``large'' charge $B$, the corresponding representation has more weight
vectors than just the Weyl images of $B$. Some of the weight vectors
will therefore have reduced length as compared to the length of $B$.
This is intenterpreted as monopole screening: smooth monopoles can
surround the 't Hooft operator inserted at the equator and screen its
magnetic charge, so that the magnetic charge $h$ observed at the polar
region is ``smaller'' than the charge $B$ of the monopole inserted.

There should be solutions to the Bogomolny equation describing monopole
screening, which are therefore labeled by $B$ and $h$ and form a finite
dimensional moduli space. Via Kronheimer's construction, such solutions
should be mapped to ASD connections on $\mathbb C^2$ which are invariant
under $\psi$-translation symmetry $U(1)_\psi$. Therefore, for $U(N)$ gauge
group the moduli space of monopoles is parametrized by the ADHM data
\[
\Big\{
 B_{1~(k\times k)}~,~
 B_{2~(k\times k)}~,~
 I_{~(k\times N)}~,~
 J_{~(N\times k)}
\Big\}~~\text{s.t.}~~
 [B_1,B_2]+IJ=0
\]
satisfying also the condition of $U(1)_\psi$ invariance.
The number $k$ and the action of $U(1)_\psi$ are determined in the
following way. Consider solutions to Bogomolny equation in which the
charge of a singular monopole $M$ is reduced to $M'$ by screening
effect. The charges $M,M'$ here are regarded as $N\times N$ diagonal
matrices. Then there should be a diagonal matrix $K$, whose size $k$
and elements are determined by the formula
\begin{equation}
 \text{Tr}(x^M) =  \text{Tr}(x^{M'}) + (x+x^{-1}-2)\,\text{Tr}(x^K)\,.
\end{equation}
Then the condition of $U(1)_\psi$ invariance is given by
\begin{equation}
 [K,B_1]+B_1\;=\;[K,B_2]-B_2\;=\;KI-IM' \;=\;M'J-JK\;=\;0.
\end{equation}
Equivariant integral on this moduli space contributes another factor to
the integrand of (\ref{HO:eq:tbg}). The detail of the analysis is
presented in \cite{HO:Gomis:2011pf} for the example of 't Hooft operators
of higher spin representations of $SU(2)$.

\subsection{Surface operators}\label{HO:sec:sfc}

Another important example of supersymmetric observables are surface
operators, which are non-local operators supported on two-dimensional
submanifolds. It will be a challenging problem to give a complete
classification of surface operators for general 4D gauge theories, but
a major progress have been made for BPS surface operators in
${\cal N}=2$ supersymmetric theories, as we review here.

For ${\cal N}=2$ theories of class S where M5-brane interpretation is
available, a natural question is how to identify the surface operators
describing other M-branes ending on or intersecting the M5-branes
\cite{HO:Alday:2009fs,HO:Alday:2010vg}. For those surface operators,
the calculations in gauge theories can be checked against the prediction
from AGT correspondence. Another approach is to realize ${\cal N}=2$
SUSY theories geometrically using Calabi-Yau compactification of type
IIA string, where D4-branes wrapping Lagrangian submanifolds give rise
to surface operators \cite{HO:Dimofte:2010tz}. In this setting, the
results of gauge theory analysis can be compared with topological string
amplitudes for which there are powerful formalisms known such as refined
topological vertex.

For ${\cal N}=2$ SUSY theories on Omega-background
$\mathbb R^4_{\epsilon_1,\epsilon_2}$ with coordinates
$z_1=x_1+ix_2, z_2=x_3+ix_4$, one can introduce surface operators along
the surfaces $z_2=0$ or $z_1=0$ without braking supersymmetry.
For theories on the ellipsoid (\ref{HO:eq:ell}), one can introduce BPS
surface operators along the $S^2$ defined by $x_3=x_4=0$ or $x_1=x_2=0$.

\paragraph{Coupled 2D-4D systems}

One way to describe surface operators is in terms of 2D quantum
field theories on its worldvolume. For 4D ${\cal N}=2$
theories realized on $S^4_b$, the objects of
interest are the half-BPS surface operators which support
${\cal N}=(2,2)$ field theories on a squashed $S^2$.
The supersymmetry for the coupled 2D-4D system is such that the $S^4_b$
and $S^2$ have the north and south poles in common, that is where the
instantons of 4D gauge theory and vortices of 2D theory get localized.

If the 4D theory has a Lagrangian description, one can simplify the
problem by turning off the 4D gauge coupling. The system is then reduced
to a 2D interacting theory and 4D free matter theory both coupled to
some frozen 4D vector multiplets. One can still learn a great deal about
surface operators from this simplified system \cite{HO:Gomis:2014eya}.
The partition function is then a product of the 4D
free hypermultiplet path integral, $Z^\text{hyp}_\text{1-loop}$ of
(\ref{HO:eq:zol}), and the $S^2$ partition function
\cite{HO:Benini:2012ui,HO:Doroud:2012xw} of the 2D theory. The classical
value of the frozen 4D vector multiplet enters the formula as the common
mass for 2D and 4D fields.

As an example, take a system of $N^2$ free hypermultiplets. One can
regard it as a bifundamental of the group $S[U(N)\times U(N)]$ and turn
on the masses $(m_1,\cdots,m_N;\tilde m_1,\cdots,\tilde m_N)$. The
$S^4_b$ partition function is then
\begin{equation}
 \prod_{i,j=1}^N\Upsilon(\tfrac Q2+i(m_i-\tilde m_j))^{-1}.
\label{HO:eq:4dh}
\end{equation}
This simple theory is known to crrespond to $N$ M5-branes wrapped on a
sphere with three (one {\it simple} and two {\it full}) punctures. AGT relation
identifies (\ref{HO:eq:4dh}) with the corresponding three-point function
in Toda conformal field theory. Now introduce a ${\cal N}=(2,2)$ theory
with the same global symmetry $S[U(N)\times U(N)]$ on the surface
operator. The simplest class of examples is a 2D $U(K)$ gauge theory with $N$
fundamental and $N$ anti-fundamental chiral multiplets. A systematic
study and detailed comparison with Toda CFT correlators were made in
\cite{HO:Gomis:2014eya}. It was shown that, if a suitable mass is turned on
for the 2D theory, which is related to $(m_i;\tilde m_i)$ by a suitable
rescaling and imaginary shift, then the 2D-4D combined partition
function reproduces the Toda four-point functions with various
degenerate insertions \cite{HO:Alday:2009fs}.

\paragraph{Singularity along a surface}

Another way to define surface operators is to require that the gauge
field and possibly other fields develop singularities along the
surface. As an example, take an $SU(N)$ gauge theory on $\mathbb C^2$ with
coordinate $z_1=x_1+ix_2$ and $z_2=x_3+ix_4$. One can then introduce a
surface operator along $z_2=0$ by imposing the singular boundary condition
\begin{equation}
 A \simeq A_\chi\cdot {\rm d}\chi\quad
 \bigg(
 \chi\equiv\arg(z_2),~~
 A_\chi\equiv\text{diag}
 (\underbrace{\nu_1,\cdots,\nu_1}_{n_1\text{ times}}\,,\,
  \underbrace{\nu_2,\cdots,\nu_2}_{n_2\text{ times}}\,,\,
  \cdots\,,\,
  \underbrace{\nu_s,\cdots,\nu_s}_{n_s\text{ times}})
 \bigg)\,.
\label{HO:eq:sgw}
\end{equation}
This breaks the gauge symmetry $SU(N)$ to a Levi subgroup
\begin{equation}
 \mathbb{L}=S[U(n_1)\times \cdots\times U(n_s)],\quad
 \sum_{i=1}^sn_i=N
\end{equation}
on the surface. The parameters $\nu_i$ satisfy
$\nu_1>\cdots>\nu_s>\nu_1-1$, which in turn set the order of
$n_i$ appearing in the partition of $N$. For half-BPS surface operators
in ${\cal N}=2$ supersymmetric theories, one needs to turn on the
auxiliary field $D_{AB}$ to ensure the SUSY variation of gaugino to
vanish. For a suitable choice of unbroken supersymmetry one finds
\begin{equation}
 D_{11}=D_{22}=0,\quad
 D_{12}=iF_{12}= 2\pi i A_\chi\cdot\delta(x_3)\delta(x_4).
\end{equation}

\vskip2mm

We have seen two different descriptions of surface operators, but there
are some surface operators described in both ways. For example, the
surface operators of type (\ref{HO:eq:sgw}) in pure ${\cal N}=2$ SYM
theory can also be described by a 2D ${\cal N}=(2,2)$ supersymmetric
quiver gauge theory which flows to a sigma model on a flag manifold
$SU(N)/\mathbb L$.
Here the ordering of $n_i$ makes a subtle effect: different orderings
leads to different ultraviolet gauge theory descriptions, which flow to
a non-linear sigma model on the same flag manifold but with different
complex structures \cite{HO:Kanno:2011fw}.

\paragraph{Localization computation}

Let us consider the surface operator of the type (\ref{HO:eq:sgw})
introduced along the $S^2$ inside the ellipsoid (\ref{HO:eq:ell}) defined
by $x_3=x_4=0$. In terms of the polar coordinates
$(\rho,\theta,\varphi,\chi)$ the surface operator is at $\theta=0$. The
singular behavior of the gauge field is then expressed as follows,
\begin{equation}
 A= A_\chi\cdot {\rm d}\chi\,. \quad (\text{near }\theta=0)
\end{equation}
At supersymmetric saddle points the gauge field takes precisely this
form. The value of classical action at the saddle points labeled by
$\phi=\bar\phi=-ia_0/2$ is
\begin{equation}
 S_\text{YM}=\frac{8\pi^2}{g^2}\text{Tr}
 \left(\ell\tilde\ell a_0^2-2i\ell a_0A_\chi \right),\quad
 S_\text{FI} = -16i\pi^2\zeta\left(\ell\tilde\ell a_0-iA_\chi\ell\right)\,,\quad
 S_\text{mat}=0\,.
\end{equation}

The saddle points can also have point-like instantons or
anti-instantons localized at the north or south poles. Due to the
presence of surface operator, the topology of gauge field configuration
near the north pole is characterized by instanton number $k$ as well as
magnetic flux $m_i$ defined by
\begin{equation}
 \frac1{2\pi} \int_\text{surface op}\text{Tr}(A_\chi\cdot F)
 ~=~ \sum_{i=1}^s\nu_im_i\,.\quad
\left(\sum_{i=1}^sm_i=0\right)
\end{equation}
Such topologically non-trivial gauge field configurations are called
ramified instantons. The saddle points with point-like ramified
instantons labeled by $k,m_i$ are thus weighted by a factor
$q^{k-\nu_im_i}$ in the path integral. Similarly, anti-instantons
localized at the south pole are labeled by $\tilde k,\tilde m_i$ and
make contributions proportional to $\bar q^{\tilde k-\nu_i\tilde m_i}$.
Those contributions are organized into a generalization of
Nekrasov's instanton partition function.

Nekrasov's partition function for ramified instantons is a generating
function of equivariant integrals over the moduli spaces
${\cal M}^\text{ram}_{k,\vec m;\vec n}$. In mathematics literature these
spaces are called Affine Laumon space. The equivariant parameters are
$\epsilon_1=\ell^{-1}$, $\epsilon_2=\tilde\ell^{-1}$ and the constant
value of the field $\hat\Phi$ at saddle points
\begin{equation}
 \hat\Phi= a_0-\frac{iA_\chi}{\tilde\ell}\,.
\end{equation}

Actually, this space ${\cal M}^\text{ram}_{k,\vec m;\vec n}$ is known
to be mathematically equivalent to another space which should be more
familiar to physicists, that is the moduli space of $U(N)$
instantons in orbifold $\mathbb C\times(\mathbb C/\mathbb Z_s)$
\cite{HO:Biswas:1997aa}. Here the $\mathbb Z_s$ is understood to act on fields
through spacetime rotation as well as gauge transformation: it acts on
the fundamental representation of $U(N)$ as the multiplication by the
diagonal matrix
\begin{equation}
 \Omega_{\vec n} \equiv \text{diag}\Big(
 \underbrace{\omega,\cdots,\omega}_{n_1}\,,\,\cdots\,,\,
 \underbrace{\omega^{s},\cdots,\omega^{s}}_{n_s}\Big)\,;\quad
 \omega\equiv e^{\frac{2\pi i}s}\,.
\label{HO:eq:omn}
\end{equation}
Each instanton is assigned a $\mathbb Z_s$ charge, and the
moduli space is denoted as ${\cal M}^\text{orb}_{\vec k;\vec n}$ with
$k_i$ the number of instantons with $\mathbb Z_s$ charge $i$
(we work with the convention $k_i=k_{s+i}$).
The two moduli spaces are related as follows,
\begin{equation}
 {\cal M}^\text{ram}_{k,\vec m;\vec n}={\cal M}^\text{orb}_{\vec k;\vec n}
 \quad\text{if}\quad
 k_s=k\,,~~ k_{i+1}=k_i+m_i\,.
\end{equation}
For more explanation, see \cite{HO:Kanno:2011fw} and references therein.

The moduli space ${\cal M}^\text{orb}_{\vec k;\vec n}$ can be
parametrized by ADHM matrices. Let us denote $K\equiv\sum_{i=1}^sk_i$,
then the set of matrices
\begin{equation}
 \left\{
 B_1{}_{(K\times K)},~B_2{}_{(K\times K)},~
 I_{(K\times N)},~
 J_{(N\times K)}
 \right\}~~\text{ s.t. }~~[B_1,B_2]+IJ=0,
\end{equation}
subject to the $\bigotimes_iGL(k_i)$ equivalence and the
$\mathbb Z_s$ orbifold projection,
\begin{equation}
\def\arraystretch{1.1}
\begin{array}{rcl}
 \Omega_{\vec k}B_1\Omega_{\vec k}^{-1}&=& B_1,\\
 \Omega_{\vec k}B_2\Omega_{\vec k}^{-1} &=& \omega B_2,\\
 \Omega_{\vec k}I\Omega_{\vec n}^{-1} &=& I,\\
 \Omega_{\vec n}J\Omega_{\vec k}^{-1} &=& \omega J.
\end{array}
\end{equation}
gives a parametrization of the moduli space
${\cal M}^\text{orb}_{\vec k;\vec n}$. Here $\Omega_{\vec k}$ is a
diagonal matrix defined similarly to (\ref{HO:eq:omn}), with eigenvalue
$\omega^i$ appearing $k_i$ times. The chain-saw quiver describes the
components of ADHM matrices which survive the orbifold projection.
\begin{center}
\includegraphics[width=9cm,bb=0 0 170 105]{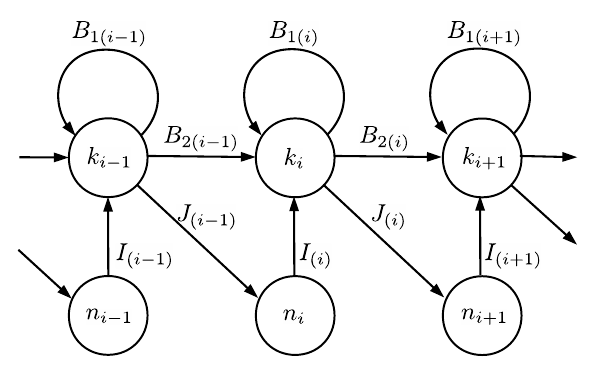}
\end{center}
Ramified instanton partition functions and their correspondence with
conformal blocks for general $W_N$ algebra were studied in
\cite{HO:Alday:2010vg,HO:Kozcaz:2010yp,HO:Awata:2010bz,HO:Wyllard:2010rp,
HO:Wyllard:2010vi,HO:Kanno:2011fw}.

The correspondence between ramified instantons and instantons in
orbifolds will be a key to fully understand how to define and compute
observables in the surface defect backgrounds. This was used in
\cite{HO:Nawata:2014nca} for surface operators in ${\cal N}=2$ pure SYM and
${\cal N}=2^\ast$ SYM theories on $S^4$, and should be extended to more
general cases. The exact formulae for observables obtained this way will
also help clarifying how various descriptions of surface operators are
related with each other.

\section*{Acknowledgments}

The author thanks Naofumi Hama for collaboration on the topics discussed
in this review. He also thanks Heng-Yu Chen and Hee-Cheol Kim for useful
discussions.

\documentfinish